\documentclass[11pt]{article}
\usepackage{graphicx}
\usepackage{subfigure}
\newcommand{\be}{\begin{equation}}
\newcommand{\ee}{\end{equation}}
\newcommand{\pr}{\prime}

\newcommand{\p}{\partial}
\newcommand{\f}{\frac}
\newcommand{\ri}{\rightarrow}

\newcommand{\bt}{\beta}

\newcommand{\D}{\displaystyle}

\begin{document}
\begin{titlepage}
\begin{flushright}
NTUA- 4/01
\end{flushright}

\vskip2truecm

\begin{center}
\begin{large}
{\bf The Phase Diagram for the anisotropic
SU(2) Adjoint Higgs Model in 5D: \\ Lattice
Evidence for Layered Structure}

\end{large}
\vskip1truecm

P.~Dimopoulos$^{(a)}$\footnote{E-mail: petros@ecm.ub.es},
K.~Farakos$^{(b)}$\footnote{E-mail: kfarakos@central.ntua.gr}, and
G. Koutsoumbas$^{(b)}$\footnote{E-mail:
kutsubas@central.ntua.gr}

\vskip1truecm

{\sl $^{(a)}$ ECM, University of Barcelona, Diagonal 647, 08028 Barcelona,
Spain} \\
{\sl  $^{(b)}$ Physics Department, National Technical University\\
15780 Zografou Campus, Athens, Greece}

\end{center}
\vskip1truecm

\begin{abstract}

\noindent We explore, by
Monte Carlo and Mean Field methods, the
five--dimensional SU(2) adjoint Higgs model.
We allow for the possibility of different couplings
along one direction, describing the so--called anisotropic model.
This study is motivated by the possibility of the appearance of
four--dimensional layered dynamics. Actually, our results lead to
the conclusion that the establishment of a layered phase in
four dimensions described by U(1) symmetry  is
possible, the extra dimension being confined due to the SU(2)
gauge symmetry. The five-dimensional
adjoint Higgs model turns out to have a layered phase, in
contradistinction with what is known about the pure $SU(2)$ model.

\end{abstract}
\end{titlepage}

\section{Introduction}

The idea of extra dimensions has been around since the time of
Kaluza and Klein \cite{kk}. The main goal has been a unified and consistent
description of the evolution of the Universe and a unification of
the known interactions. The approach is appealing, although no
experimental or observational facts give any compelling
evidence to support such a hypothesis. One of the main problems
of the method is the connection with four-dimensional Physics.

During the last four years there has been a revived interest in such
models, mainly due to the concern about the hierarchy problem.
There are two versions of using extra dimensions, depending on whether
they are considered as compactified \cite{sdimo} or not \cite{rs}.
Much work have been already done on
the graviton localization problem in the four-dimensional
space (3-brane) which is embedded in a higher dimensional space
and furthermore on problems related with the localization of
gauge and matter fields \cite{alex}.

Fu and Nielsen in 1984 \cite{funiel} originated work on
the problem of a higher--dimensional space on the lattice. They
considered  a five dimensional pure U(1) gauge theory and showed,
by Mean Field methods, that for certain values of the
couplings for the $n$ extra dimensions four dimensional layers may
be formed within the $(4+n)$-dimensional space.
The main feature  of the layer space is Coulomb interaction
along the layer combined with confinement
along the extra dimensions. Actually, it is this confinement
which forbids interaction with neighbouring layers and manifests our
effective detection of a four dimensional world. Many works on the lattice
have been appeared and are based on these ideas, using either a pure $U(1)$
theory or the Abelian Higgs Model.
The lattice results (\cite{stam}, \cite{dim1})
have given serious evidence for a layered structure in either
the Coulomb or the Higgs phase.

It should also be mentioned that consideration of non-Abelian
gauge theories in relation with the layered structure has been
already performed both at zero \cite{rab} and finite
temperature \cite{china}.
Especially, for the case of zero temperature there is  a
prediction that the layered structure in pure non-Abelian theories
may exist if the theory is defined in six or higher dimensions.
Also  the pure SU(2) theory in five
dimensions with anisotropic couplings has been studied in
\cite{kubo} but in a different context since the theory is studied
in the region where $\beta_{g}^{\prime} \gg \beta_{g}$, the
lattice spacing in the fifth direction being smaller than the
corresponding ones along the other four directions to simulate a
small compactification radius.

In this paper we consider the five--dimensional SU(2)--adjoint
Higgs model with anisotropic couplings on the lattice \footnote{
For the  analysis of the three and four dimensional SU(2)--adjoint
Higgs model see
\cite{nad,MZ}.}.
Actually, we explore the phase
diagram of the model and we find that for certain values of the
couplings a layered structure in Higgs phase does really exist and that
it is associated with the unbroken U(1) symmetry in the Higgs
phase. Before introducing the details concerning the phases of the
model it should be mentioned that although the pure SU(2) theory
exhibits a non--physical five--dimensional layered phase, the
introduction of the scalars in the adjoint representation
opens up the possibility of the appearance of the layered phase
in four dimensions. Furthermore, as one
can see from our results, the confinement along the extra
dimension is due to a SU(2) strong interaction which gives a
possible hint for the physical implications of the
model in accordance with the idea of the four-dimensional layer
formation. This procedure may be connected with some old efforts
which proposed the localization of the fields on a membrane
playing the role of our four--dimensional world which is embedded
in a higher-dimensional one \cite{Barnaveli}. The idea of the SU(2)
confinement in the transverse dimension with subsequent localization of the
fields on the remaining ones can be found in \cite{Dvali}, although
this work was refering to a four--dimensional space--time.

In section 2 we write down the lattice action for the model and we
present the order parameters which we have studied.
Monte Carlo results are presented in
section 3; in particular we present results on both the isotropic
and the anisotropic models.
In the Appendix we present the Mean Field Analysis of the
model from which many interesting conclusions can be derived about
the behaviour of the system. Also the comparison of the Mean Field
against the Monte Carlo results gives a feeling about the validity of
the Mean Field approach, which is useful if one wishes to use it
in situations where the Monte Carlo simulation is too time
consuming.

\section{The Lattice action and the order parameters}
In order to express the anisotropic SU(2)--adjoint Higgs model on
the lattice we single out the direction $ \hat{5}$ by couplings
which differ from the corresponding ones in the remaining four
directions. Thus, we write down explicitly:
\begin{eqnarray}
S=& &\beta_{g} \sum_{x, 1 \leq \mu \leq \nu} (1-\frac{1}{2} Tr U_{\mu \nu}(x)) +
\beta^{\prime}_{g} \sum_{x,\mu} (1-\frac{1}{2} Tr U_{\mu 5}(x))
\nonumber \\
&+& \beta_{h} \sum_{x, \mu} \left(\frac{1}{2}
Tr\left[\Phi^{2}(x)\right]
- \frac{1}{2} Tr\left[\Phi(x)U_{\mu}(x)
\Phi(x+\hat{\mu})U^{\dag}_{\mu}(x)\right]\right) \nonumber \\
&+& \beta^{\prime}_{h} \sum_{x}
\left(\frac{1}{2} Tr\left[\Phi^{2}(x)\right] - \frac{1}{2} Tr\left[\Phi(x)U_{5}(x)
\Phi(x+\hat{5})U^{\dag}_{5}(x)\right]\right) \nonumber \\
&+& (1-2 \beta_{R} -4 \beta_{h} -\beta^{\prime}_{h}) \sum_{x} \frac{1}{2}
 Tr\left[\Phi^{2}(x)\right] +  \beta_{R} \sum_{x} \left(\frac{1}{2}
 Tr\left[\Phi^{2}(x)\right]\right)^{2}
 \label{action}
\end{eqnarray}
where $U_{\mu}=e^{igA_{\mu}}$ and  $U_{5}=e^{igA_{5}}$.
$U_{\mu \nu}$ and $U_{\mu 5}$ are
the plaquettes defined on the four dimensional space and along the
extra direction respectively.
The gauge potential and the matter fields are represented by
the $2 \times 2$ Hermitian matrices $A_\mu=A_\mu^a \sigma_{a}$ and
$\Phi=\Phi^{a} \sigma_{a}$ respectively,
where $\sigma_{a}$ are the Pauli matrices.
The couplings refering to the fifth
direction are primed to distinguish them from the ``space--like"
ones.

The order parameters that we use are separated in space--like and
transverse--like ones, except from the Higgs field measure squared
which does not depend on the direction. The relevant definitions follow:

\be
{\rm Space-like~Plaquette:~~~} P_S \equiv <\f{1}{6 N^5} \sum_x
\sum_{1 \le \mu<\nu \le 4} Tr U_{\mu \nu}(x)>
\ee
\be
{\rm Transverse-like~Plaquette:~~~} P_T \equiv <\f{1}{4 N^5}
\sum_x \sum_{1 \le \mu \le 4} Tr U_{\mu 5}(x)>
\ee
\be
{\rm Space-like~Link:~~~} L_S \equiv <\f{1}{4 N^5} \sum_x
\sum_{1 \le \mu \le 4} \frac{1}{2} Tr\left[\Phi(x)U_{\mu}(x)
\Phi(x+\hat{\mu})U^{\dag}_{\mu}(x)\right] / \frac{1}{2} Tr\left[\Phi^{2}(x)\right] >
\ee
\be
{\rm Transverse-like~Link:~~~} L_T \equiv <\f{1}{N^5} \sum_x
\frac{1}{2} Tr\left[\Phi(x)U_{5}(x)
\Phi(x+\hat{5})U^{\dag}_{5}(x)\right] / \frac{1}{2} Tr\left[\Phi^{2}(x)\right]>
\ee
\be
{\rm Higgs~field~measure~squared:~~~} R^2 \equiv \f{1}{N^5} \sum_x
\frac{1}{2} Tr\left[\Phi^{2}(x)\right]
\ee

In the above equations $N$ is the linear dimension of
the symmetric $N^5$ lattice.

The behaviours of each of the chosen order parameters which
characterize the various phases of the system
will be explained in the next sections.

\section{Monte Carlo Results}
In performing the simulations we used the
Kennedy--Pendleton heat bath algorithm for the updating of the gauge field
and the Metropolis algorithm for the Higgs field.  The
exploration of the phase
diagram was done using the hysteresis loop
method. Except for the cases where large
hysteresis loops are present and indicate first order phase transitions,
our analysis is not accurate enough to determine the order of
the phase transitions, in particular of the continuous ones.
The phase transition lines
have been determined up to the accuracy provided by the hysteresis
loop method. The simulations have been performed mainly on
$4^5$ and $6^5$ lattice volumes. However at selected phase space
points we have used $8^5$ lattice volumes to better determine
the location of the phase transition lines.

We first present the phase diagram for the isotropic model and
then we go on with the anisotropic one for two values of the
$\beta_{g}$ lattice gauge coupling, namely $\beta_{g}=0.5 \
\mbox{and} \ 1.2$. The reason for these choices will become clear soon
and is related with the possible gauge symmetry surviving in the
layered phase.

\subsection{The Isotropic Model}

When we consider the isotropic model,
the action is given by (\ref{action}) where
$\beta_g^\pr=\beta_g$ and $\beta_h^\pr=\beta_h. $
Before proceeding we should state some analytical predictions which
are extremely useful
for the characterization of the various phases of our model and
will be heavily used in the sequel.
These predictions concern the values of the plaquette in the strong
and weak coupling phases for a pure gauge lattice theory.
We expect
that the original $SU(2)$ symmetry of the model will eventually break
down to $U(1),$ since the scalar field belongs to the
adjoint representation.
Thus we present the relevant results for
both symmetry groups. It is convenient to present these predictions for the
isotropic case, since the values of the space-like and the time-like
plaquettes coincide $(P_T=P_S\equiv P).$
Let us first consider $SU(2).$ We know
that for a strongly coupled model $(\beta_g << 1)$ the plaquette
is given by $$P \approx  \D{\f{\beta_g}{4}},\ {\rm if} \ \beta_g << 1. $$
In the weak coupling for a $D-$
dimensional $SU(2)$ theory the plaquette is approximated by: $$P \approx
1-\f{3}{D \beta_g}, \ {\rm if} \ \beta_g >> 1.$$
For $U(1)-$symmetric theories the corresponding
approximations read:
$$P \approx \f{\beta_g}{2}, \ {\rm if} \ \beta_g << 1$$ and
$$P \approx 1-\f{1}{D \beta_g},\ {\rm if} \ \beta_g >> 1.$$
We have not specified the dimension $D$ of space-time on purpose. Of
course, in this subsection, $D=5.$ However, when we consider the
anisotropic models, we may have eventually a dimensional reduction from
$D=5$ to $D=4.$

Before even performing the simulations we know some general characteristics of the
phase diagram. For $\beta_h=0$ the scalar fields decouple from the
dynamics and the model reduces to five-dimensional pure $SU(2).$ The model is
known \cite{creu, kawai} to undergo a phase transition at $\beta_g \simeq 1.63$ from
the strong coupling phase $S,$ where the plaquette behaves as
$\D{\f{\beta_g}{4}},$ to a Coulomb phase $C,$ where the plaquettes
asymptotically behave as $1-\D{\f{3}{5 \beta_g}}.$ Of course, this
critical point will be part of a phase transition line, which will
continue in the interior of the phase diagram. We also know the
characteristics of the $\beta_h \rightarrow \infty$ limit. To begin with,
the model is in the Higgs phase. Only a
$U(1)$ subgroup of the $SU(2)$ gauge symmetry will survive in this limit, so
the model will effectively be a five-dimensional
pure $U(1)$ gauge model with coupling
$\beta_g.$ This model also has a phase transition (as one varies $\bt_g)$
from the strong coupling Higgs phase $H_S$
$\left(P \approx \D{\f{\beta_g}{2}}\right)$ to
a weak coupling Higgs phase $H_C$ $\left(P \approx 1-\D{\f{1}{5 \beta_g}}\right).$

\begin{figure}[!h]
\begin{center}
\includegraphics[scale=0.45]{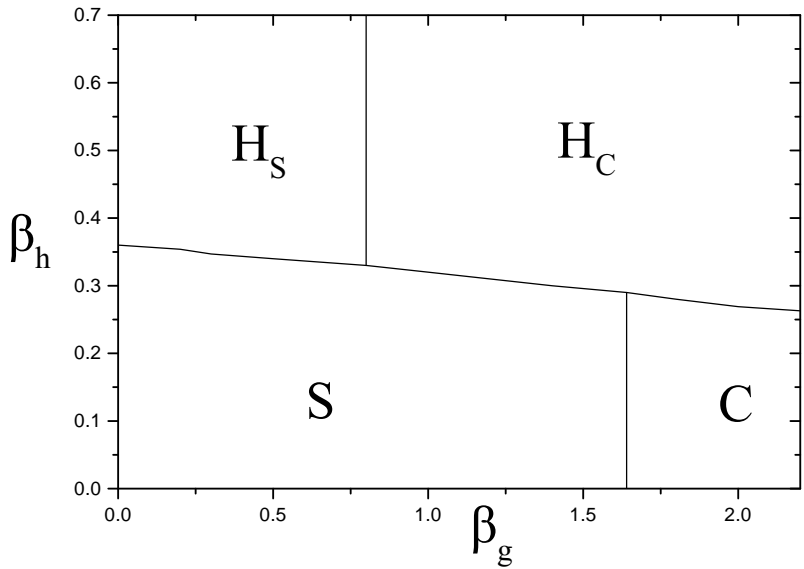}
\caption{Phase diagram for the isotropic model and for $\beta_{R}=0.01.$
With the help of the hysteresis loops all the phase transition lines
turn out to be first order. }\label{Fig1.1}
\end{center}
\end{figure}

In figure \ref{Fig1.1} we give the phase diagram for the isotropic five-dimensional
SU(2) adjoint Higgs model in terms of the $\beta_{g}$ and
$\beta_{h}$ lattice
couplings, having fixed the Higgs self--coupling constant to the value
$\beta_{R}=0.01$.
We can begin by the understanding that the value of $\bt_R$ is of
secondary importance: a small value for $\bt_R$ corresponds to strong phase
transitions, while larger values cause the various
transitions to weaken. This has been the invariable result of very
many simulations in the past for various symmetry groups, mainly
for tree- and four-dimensional space-time. We expect that something
similar will characterize our model, so we set $\bt_R=0.01$ for all
our simulations.

\begin{figure}
\subfigure[]{\includegraphics[scale=0.25]{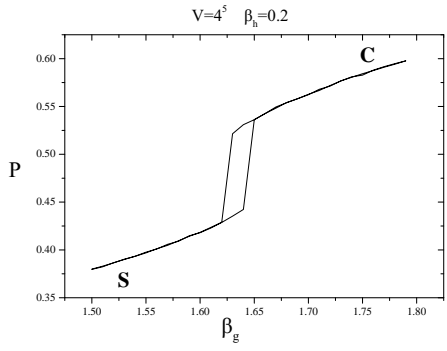}}
\subfigure[]{\includegraphics[scale=0.25]{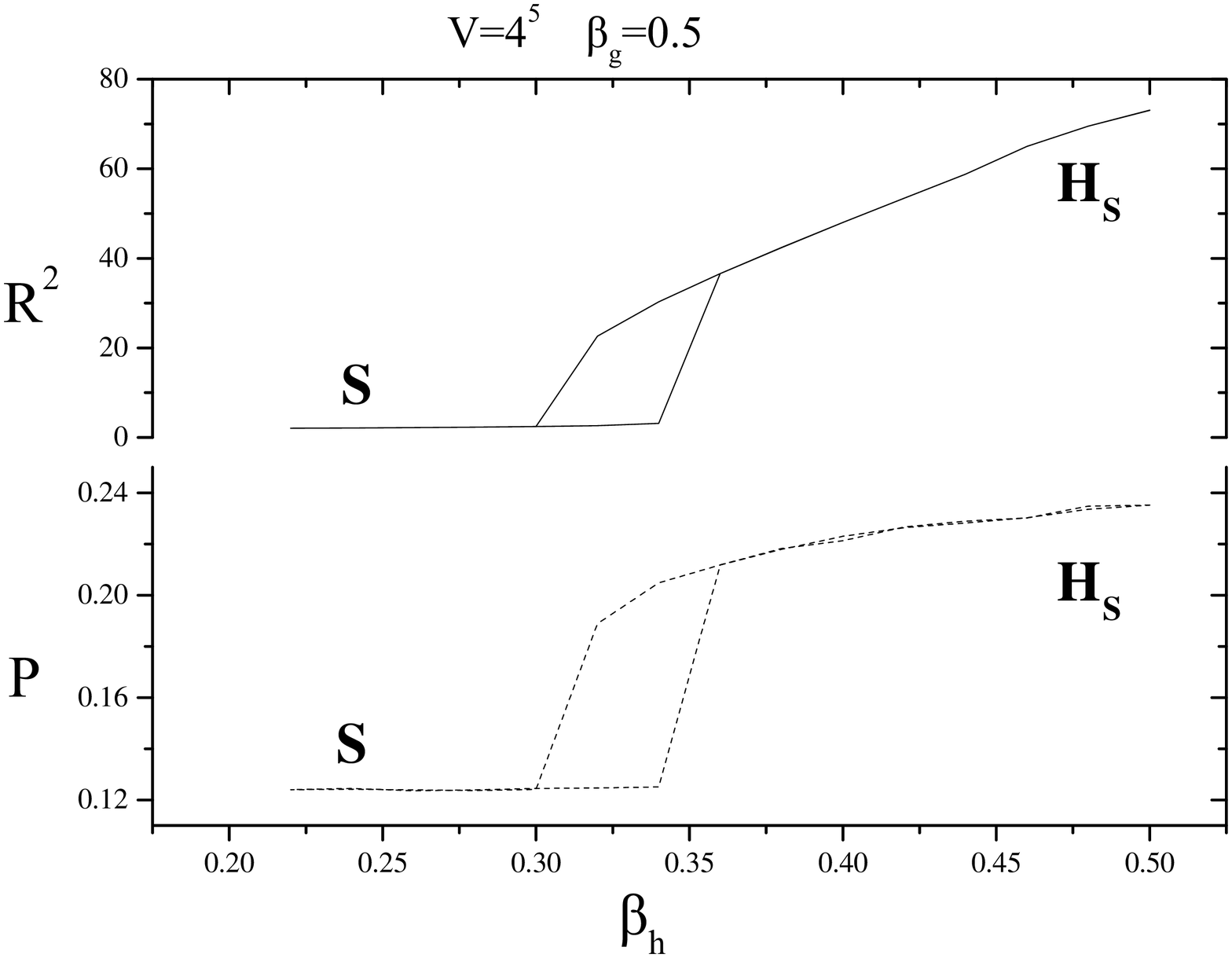}}
\subfigure[]{\includegraphics[scale=0.25]{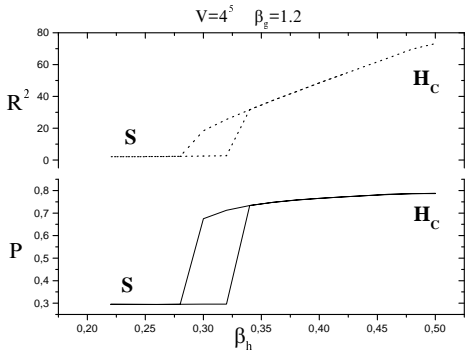}}
\subfigure[]{\includegraphics[scale=0.25]{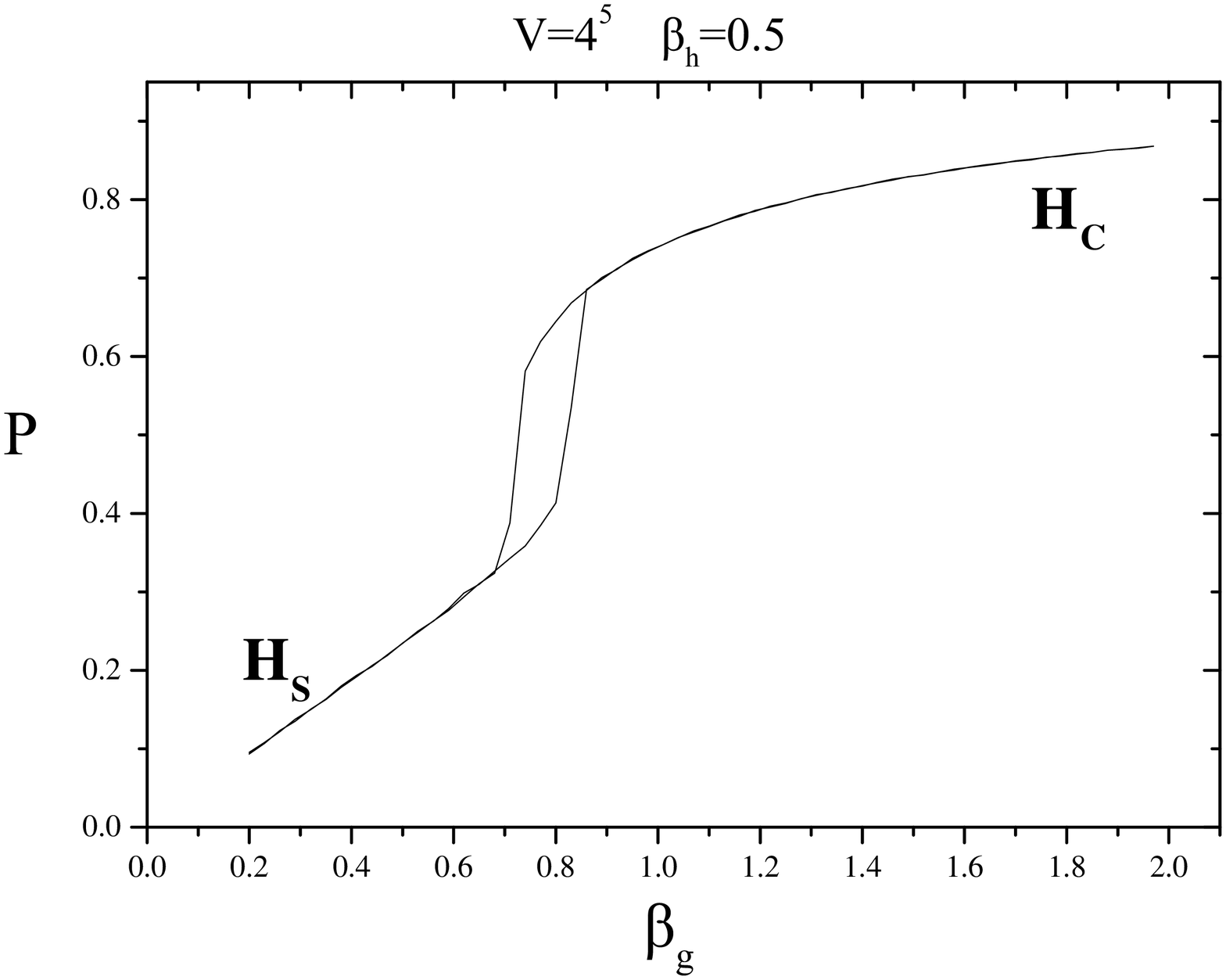}}
\subfigure[]{\includegraphics[scale=0.25]{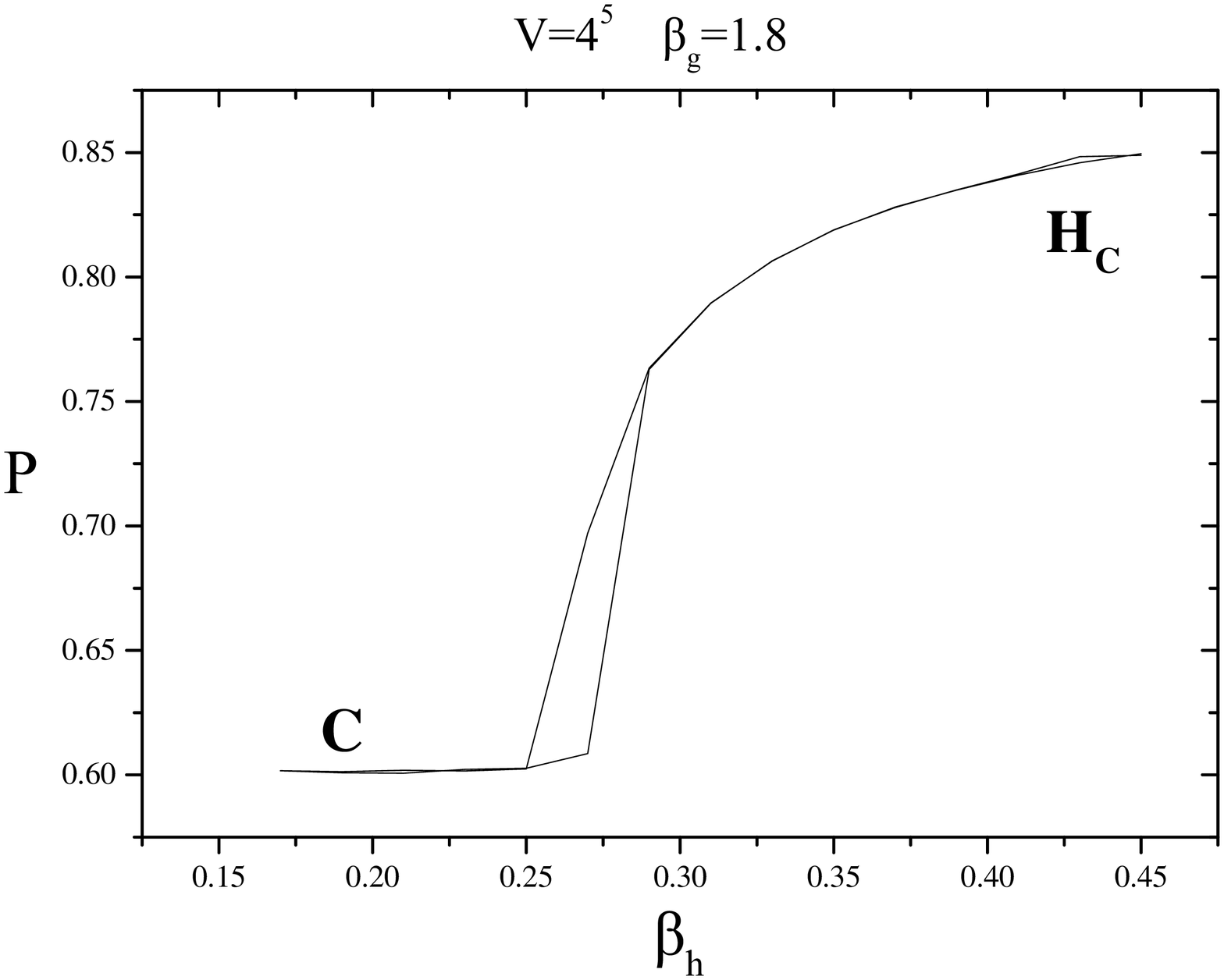}}
\caption{Sample runs on $4^5$ lattice volumes.
(a) An example for the $S-C$ phase transition where the plaquette
is depicted, for $\beta_{h}=0.2$.
(b) The $S-H_{S}$ phase transition for $\beta_{g}=0.5$ where
the plaquette and $R^2$ behaviour are shown. (c) The $S-H_{C}$ phase
transition for $\beta_{g}=1.2$
(d) The $H_{S}-H_{C}$
phase transition for $\beta_{h}=0.5$ for the plaquette values showing clearly
the transition from the Confining to Coulomb phase of U(1) gauge symmetry.
(e) The plaquette hysteresis loop for
the $C-H_{C}$ phase transition at $\beta_{g}=1.8$.}\label{Fig2.1}
\end{figure}

In figure \ref{Fig1.1}
we may also see the limiting cases that we described previously
and how they form part of
the full phase diagram. Within $S$ and $C$ we are in the phase with
unbroken symmetry, where $R^2 \sim 1.$ The phases $H_S$ and $H_C$
are characterized by a value of $R^2$ substantially bigger than $1.$
The phase transitions appear to be first order, since they exhibit
large hysteresis loops.

In the sequel (figure 2) we will present some of the results
that we used to derive the phase diagram.
In figure 2(a) we depict the plaquette as a function of $\bt_g$ for
$\bt_h=0.2.$ This line in the parameter space crosses the $S-C$
phase transition line. One may check that the expression $\D{\f{\bt_g}{4}}$
slightly underestimates the plaquette (but it is quite close). The
deviation is bigger for the Coulomb phase. We should remark here that
the analytical expression for the weak coupling
systematically overestimates the result.

Figure 2(b) contains results concerning the $S-H_S$ phase transition.
$\bt_g$ is set to 0.5 and $\bt_h$ runs.
We see the rise of $R^2$ as $\bt_h$ crosses the phase transition line,
signaling the onset of the Higgs phase. A very large hysteresis loop
appears, indicating a first order transition. The plaquette
value yields some hints about the symmetry group
prevailing in each phase. For
small values of $\bt_h$ the plaquette equals $\D{\f{\bt_g}{4}}=0.125$ with
very good precision, so it corresponds to the strong coupling phase
of a pure $SU(2)$ model. For large values of $\bt_h$ the remaining symmetry
is presumably $U(1).$ This is indeed consistent with the value of the
plaquette, which, for large enough $\bt_h,$ tends to the value
$\D{\f{\bt_g}{2}}=0.250,$ which is the signature of the strong coupling
phase of $U(1).$ Consequently, we are in the strong coupling regime of the
$U(1)$ which survives the Higgs transition.

Figure 2(c) contains material on the $S-H_C$ transition. Here $\bt_g$
is set to 1.2. Again we
check from the value of $R^2$ that we move to a Higgs phase; the
phase transition takes place at about $\bt_h = 0.30.$ Before the transition
the plaquette equals $\D{\f{\bt_g}{4}}=0.30$ to a very good precision, so
we start again with the $SU(2)$ strong coupling phase. After the
phase transition the plaquette tends to the value $1-\D{\f{1}{5 \bt_g}} = 0.83,$
which characterizes the weak coupling phase of five-dimensional $U(1).$

In figure 2(d) we depict the transition from $H_S$ to $H_C:$ $\bt_h$ is set
to a large value, $\bt_h=0.5,$ to make sure that we are in the Higgs phase,
and $\bt_g$ runs.
The plaquette in $H_S$ behaves as $\D{\f{\bt_g}{2}}$ (strong $U(1)),$
while after the transition it follows the curve $1-\D{\f{1}{5 \bt_g}}.$

Finally, figure 2(e) contains information on the $C-H_C$ phase transition:
$\bt_g$ is set to the large value 1.8 and $\bt_h$ varies. We
start with the value 0.61 for the plaquette, which is well approximated by
the formula $1-\D{\f{3}{5 \bt_g}}.$
Anyway this value characterizes the weak $SU(2).$
After the transition the plaquette tends to the value $1-\D{\f{1}{5 \bt_g}}=0.89,$
that is to a weak $U(1)$ theory.

The hysteresis loops are quite sizeable and they indicate
that the phase transitions are of first order.

\subsection{The Anisotropic Model}

The phase diagram of the anisotropic model is expected to be much
richer than the one of the isotropic model, studied in the previous
paragraph. The parameter space is very large (consisting of the
parameters $\bt_g, \bt_g^\pr, \bt_h, \bt^\pr_h, \bt_R),$ so a crucial
step is to decide which subspace is interesting to investigate.
Of course we set $\bt_R=0.01$ as before.
On the other hand we know the behaviour of the system
in the four-dimensional limit. If $\bt_g^\pr=0$ and $\bt_h^\pr=0,$
the system becomes strictly four-dimensional. Here also, the
qualitative features of the phase diagram are well known since a long
time \cite{MZ}: The four-dimensional model
will have a strong-coupling phase $S,$ for relatively small
values of $\bt_h,$ and a four-dimensional Higgs phase, $H_4,$ for
large values of $\bt_h.$ The $U(1)$ symmetry which survives in the Higgs
phase $H_4$ has a phase transition separating a strong from a weak phase.
The problem is what will happen when the
``primed" quantities $\bt_g^\pr$ and $\bt_h^\pr$ take on
non-zero values. We will first determine the phase diagram on the
$\bt_g^\pr-\bt_h$ plane, for $\bt_h^\pr=0.001$ at selected values of
$\bt_g.$ Later on, we will study the role of $\bt_h^\pr.$

We now choose the values of $\bt_g$ for which we will scan the
$\bt_g^\pr-\bt_h$ plane. The first value will be $\bt_g=0.5,$
since we would like to have a value of $\bt_g$ lying
in the strong coupling regime.
Note that (referring to figure 1),
in the isotropic model, if one starts at $\bt_g=0.5,$ for
large enough $\bt_h$ one will end up into the strong coupling Higgs
phase, $H_S.$
An interesting remark is that the confinement scale,
which is contained naturally in the theory at $\bt_g=0.5,$ may serve as
a cut-off; the cut-off is necessary to define a non-renormalizable
model, such as this one.
For completeness we also choose the value $\bt_g=1.2,$
for our second set of measurements. The difference from the
$\bt_g=0.5$ case is that, in the isotropic model, one ends up with
the weakly coupled Higgs phase $(H_C),$ if $\bt_h$ is set to a large value.
We do not try any
value of $\bt_g,$ which corresponds to the five-dimensional
Coulomb phase of $SU(2),$ that is $\bt_g>1.63,$
since starting from a weakly coupled
$SU(2)$ it is unlikely that one may end up with a layered phase,
which is the main subject of the present work. Thus we will only
study the values $\bt_g=0.5$ and $\bt_g=1.2.$

It is interesting to point out that the criterion that we used in the
previous paragraph to discriminate between phases, namely the
behaviour of the plaquettes for strong and weak couplings needs to be
modified. This is necessary, since, for instance, $\bt_g$ may lie in the weak
coupling range, while $\bt_g^\pr$ may be strong. In addition we may
encounter space-times with effective dimensions 4 or 5, so the
expressions concerning the weak coupling will also change. A first
guess at the behaviour of the plaquettes is that $P_S$ will behave
as $\D{\f{\bt_g}{4}}$ in the strong $SU(2)$ regime and as $1-\D{\f{3}{D \bt_g}}$
in the weak coupling $SU(2)$ for D dimensions. Correspondingly,
$P_T$ will be
given by similar expressions, with $\bt_g$ replaced by $\bt_g^\pr.$
The behaviours of the plaquettes if the surviving symmetry is $U(1)$
read $\D{\f{\bt_g}{2}}, \ 1-\D{\f{1}{D \bt_g}}$ for $P_S$ and
$\D{\f{\bt^\pr_g}{2}}, \ 1-\D{\f{1}{D \bt_g^\pr}}$ for $P_T.$
It turns out that these expressions are not very good approximations.
They give the qualitative flavour of each phase but are
less accurate than the corresponding formulae of the isotropic model.
We now proceed with the study of the two chosen values of $\bt_g.$

\vspace{0.5cm}

$\bullet$ \hspace{0.1cm} $\mathbf{ \beta_{g}=0.5}$
\vspace*{0.2cm}

\begin{figure}[!h]
\begin{center}
\includegraphics[scale=0.45]{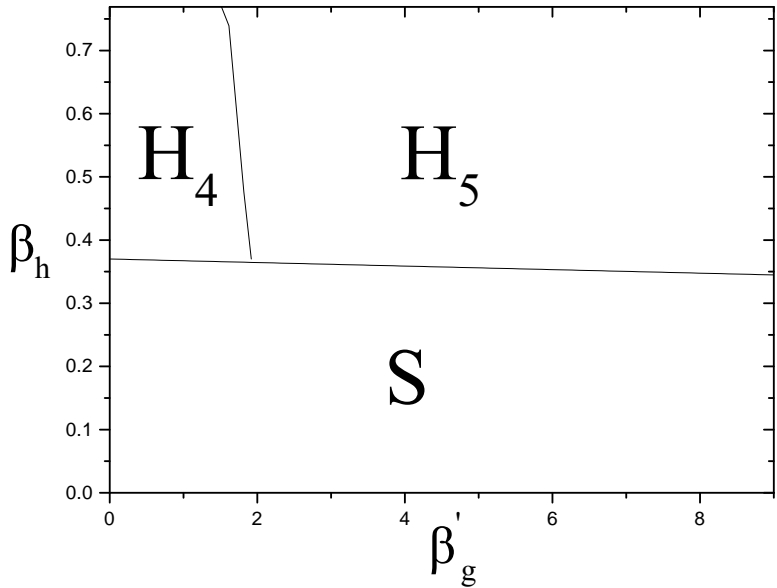}
\caption{Phase diagram for the anisotropic model for $\beta_{g}=0.5$ and
$\beta_{R}=0.01.$
Using the hysteresis loops all the phase transition lines
turn out to be first order. }\label{Fig1.2}
\end{center}
\end{figure}

We begin by giving the phase diagram that we obtain, as we did
with the isotropic model. Later on we will give the results of
some of the runs that we used to derive the phase diagram.
The phase diagram is depicted in
figure \ref{Fig1.2} and has three phases: The Strong, confining phase $S,$
the ``layered", four-dimensional Higgs phase, $H_4,$ and the
five-dimensional Higgs phase, $H_5.$ Let us explain briefly what
we understand about these phases. The $S$ and $H_5$ phases are
fairly easy to understand: they are the five-dimensional
phases with unbroken and
broken symmetry respectively. The expected
behaviours of the plaquettes are given in table 1. In particular,
in the confining phase $S,$ $P_S$ and $P_T$ will behave as
$\D{\f{\beta_{g}}{4}}$ and $\D{\f{\beta^{\prime}_{g}}{4}}$ respectively.
In the $H_5$ phase the $SU(2)$ symmetry will be broken down to a
(weak coupling) $U(1),$ so the behaviours of $P_S$ and $P_T$ are
$1-\D{\f{1}{5 \beta_{g}}}$ and $1-\D{\f{1}{5 \beta^{\prime}_{g}}}$
respectively. The $H_4$ phase needs some clarifications. When
$\bt_g^\pr=0, \bt_h^\pr=0,$ we expect that the system will be
strictly four-dimensional. When $\bt_g^\pr$ and $\bt_h^\pr$ are
small (but non-zero),
we expect that we may have some kind of layer structure
(that is, almost four-dimensional), which is characterized by
small transverse-like quantities and large space-like ones. This
is the phase that we have denoted by $H_4.$ However, in this case,
it is not only the transverse direction which is confining $\left(P_{T}
\sim \D{\f{\beta^{\prime}_{g}}{4}}\right),$ but also the space-like
behaviour is dominated by the strong coupling
$\bt_g=0.5,$ that is $P_S \sim \D{\f{\beta_g}{2}}.$ In other words, we
have a strong coupling confining behaviour both in the transverse
direction (where the $SU(2)$ symmetry remains intact, and the
denominator is $4)$ and within the layer, where the symmetry is
broken down to $U(1),$ which is the reason behind the denominator
$2.$ Thus in this phase we have $SU(2)$ confinement along
the extra fifth direction, and $U(1)$ confinement
within the layers. Since  by the term layered phase we mean a
situation with confinement in the transverse direction and Coulomb
behaviour within the layers, we remark here that this term is not
proper here, so we should enclose the word {\it layered} in quotes in
this case. Let us note that the two Higgs phases are characterised
primarily by large values for the space--like link $L_S$ and the
measure squared $R^2$. An
important remark is that within $H_4$ the $SU(2)$ symmetry does
not break down to $U(1)$ in the transverse direction. It is only
within the layers that the symmetry group changes.

\begin{center}
\begin{table}[!b]
\hspace*{2cm}
\begin{tabular}{|c|c|c|} \hline & \bf{Layer} &
\bf{Transverse direction} \\ \hline \bf{S}& SU(2)--strong: $P_{S}
\sim \D{\f{\beta_{g}}{4}}$ & SU(2)--strong: $P_{T} \sim
\D{\f{\beta^\prime_g}{4}}$ \\ \hline $\mathbf{H_{4}}$ & U(1)--strong:
$P_S \sim \D{\f{\beta_g}{2}}$ & SU(2)--strong: $P_{T} \sim
\D{\f{\beta^{\prime}_{g}}{4}}$ \\ \hline $\mathbf{H_{5}}$ & U(1)--5D
Coulomb: $P_{S} \sim 1-\D{\f{1}{5 \beta_{g}}}$ & U(1)--5D Coulomb:
$P_{T} \sim 1-\D{\f{1}{5 \beta^{\prime}_{g}}}$  \\ \hline
\end{tabular}
\caption{Phase characterization in terms of  $P_{S}$ and $P_{T}$
values for $\beta_{g}=0.5$.}
\end{table}
\end{center}

\begin{figure}[!h]
\subfigure[]{\includegraphics[scale=0.25]{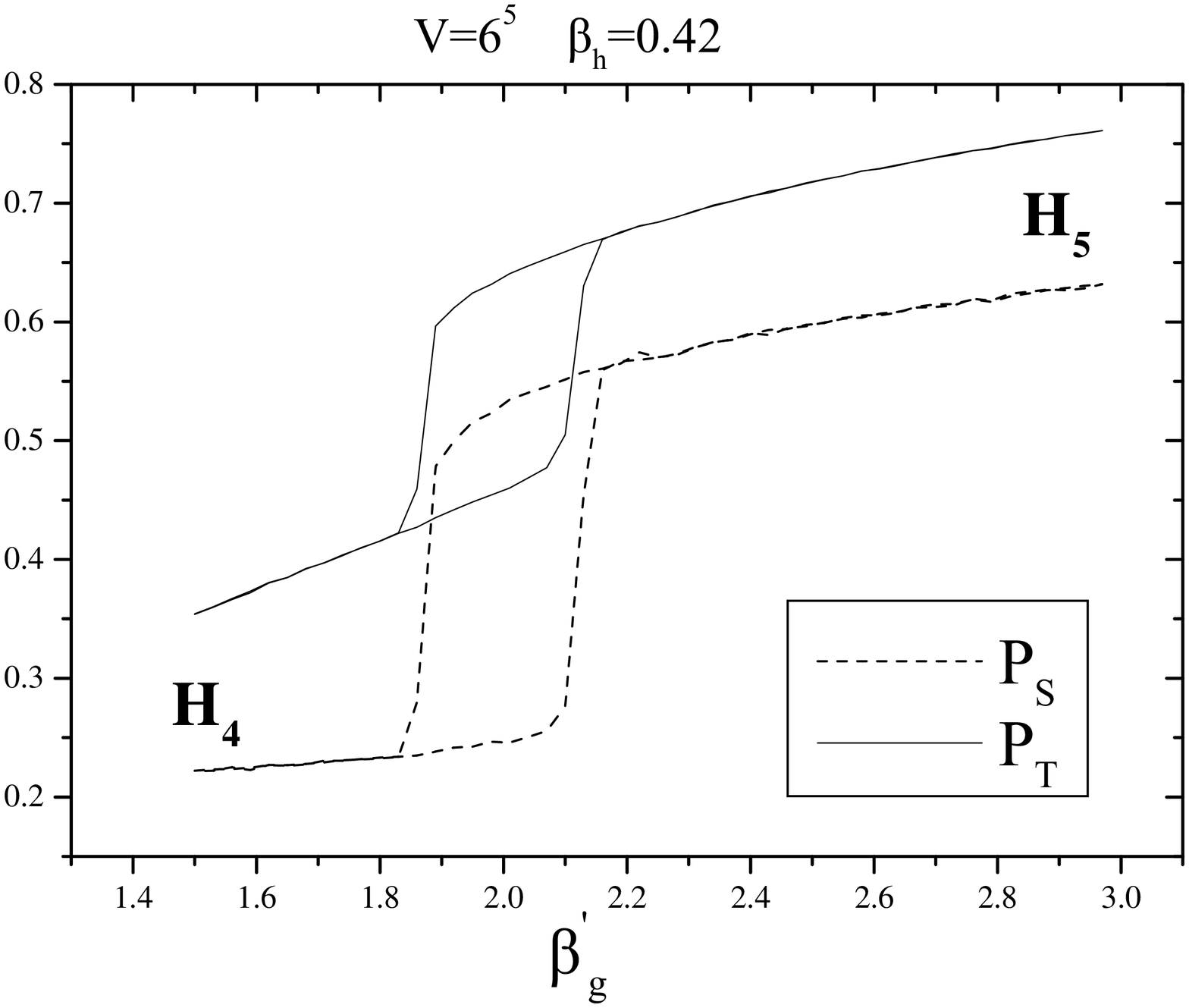}}
\subfigure[]{\includegraphics[scale=0.25]{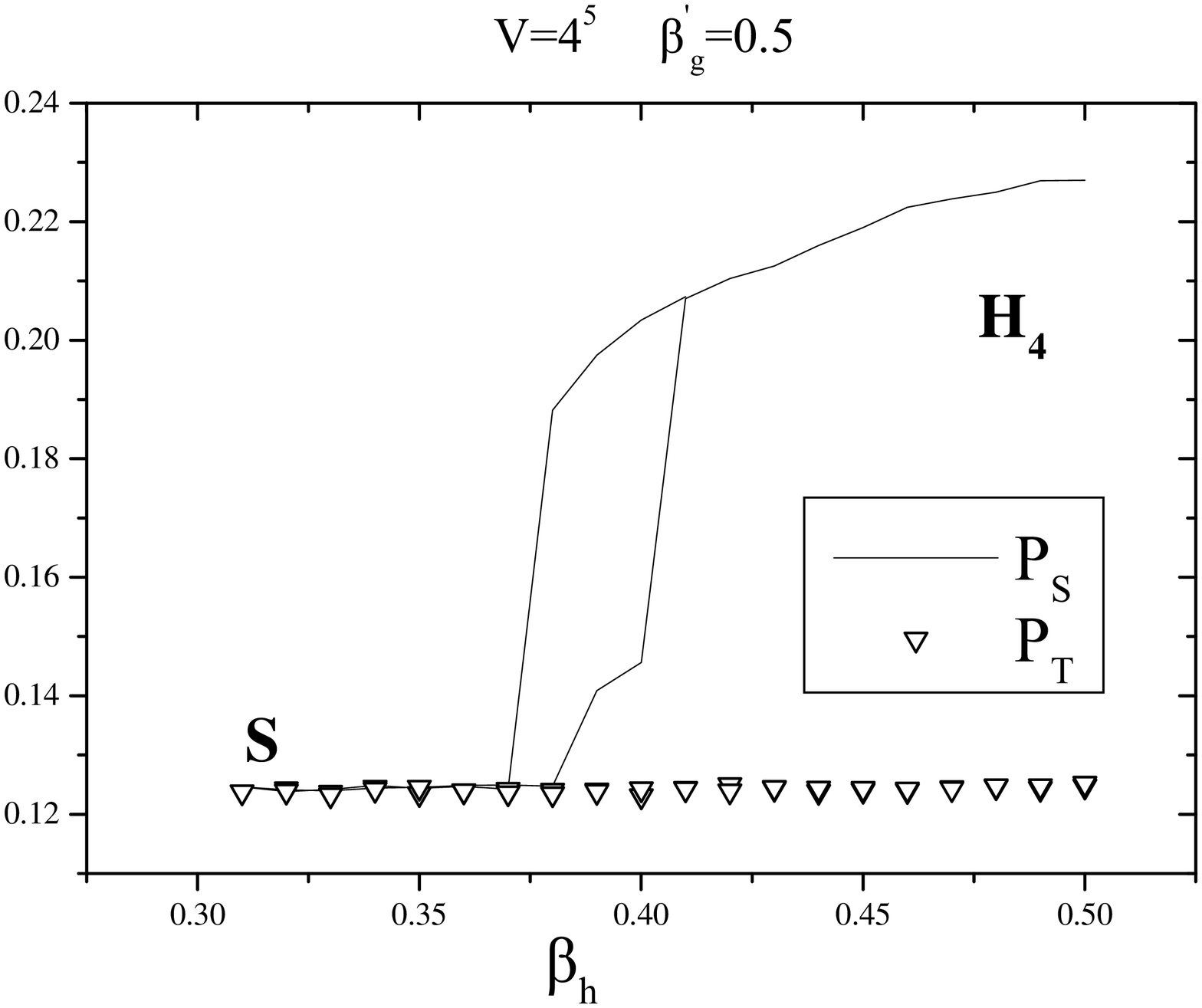}}
\subfigure[]{\includegraphics[scale=0.25]{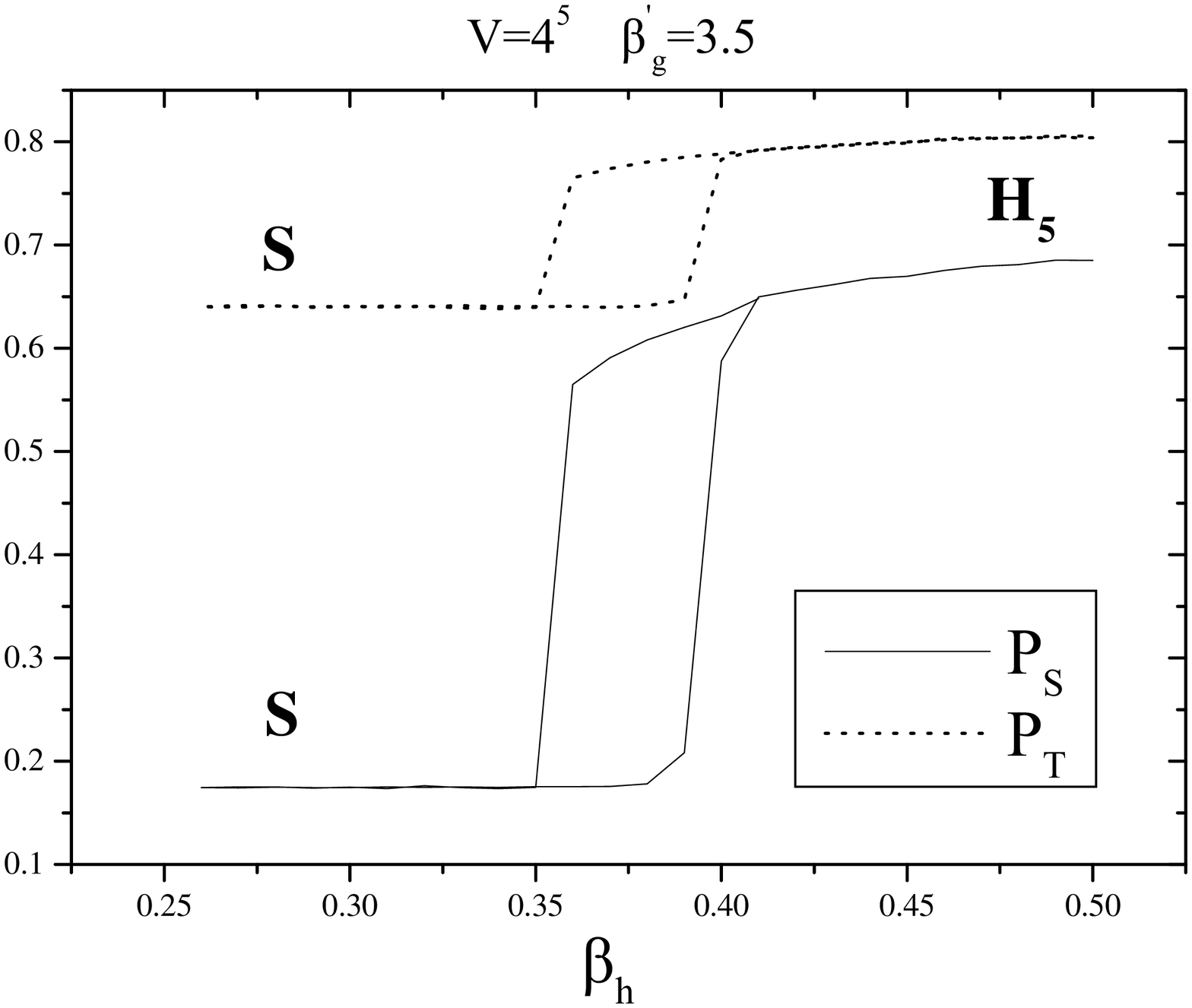}}
\caption{Examples concerning the anisotropic model for $\beta_{g}=0.5$.
(a) The $H_{4}-H_{5}$ phase transition.
(b) The $S-H_{4}$ phase transition. (c) The $S-H_{5}$ phase transition. }
\label{Fig2.2}
\end{figure}

Let us examine in some more detail the behaviour of the various
quantities across the phase transitions.

In figure 4(a) we have set $\bt_h=0.42,$ so that the system is in the Higgs
phase, and we let $\bt_g^\pr$ run. Considering $P_T$ we see that it
initially equals $\D{\f{\beta_g^\pr}{4}},$ which signals
$SU(2)$ confinement, while after the transition it tends asymptotically
to the value $1-\D{\f{1}{5 \bt_g^\pr}},$ that is to a
Coulomb phase of $U(1).$ Thus we observe
the breaking of the symmetry group in the transverse direction.
On the other side of the transition $P_S$ starts with a value quite close to
$\D{\f{\beta_g}{2}}=0.250,$ while after the phase transition it
approaches asymptotically the value $1-\D{\f{1}{5 \bt_g}}=0.60.$
Thus within the layers we have a transition from $U(1)$ confinement
to a $U(1)$ Coulomb phase.
The above findings are consistent with a transition from $H_4$ to $H_5.$

The changes across the
$S-H_4$ transition are shown in figure 4(b). $\bt_g^\pr$ has been set
to 0.5 (equal to $\bt_g)$ and the $\bt_h$ coupling runs.
It turns out that $P_T$ is
equal to $\D{\f{\beta_g^\prime}{4}}=0.125$ in both phases.
On the contrary, $P_S$ equals
$\D{\f{\beta_g}{4}}=0.125$ (equal to $P_T)$ in the strong coupling phase,
while after the transition it approaches (for large enough $\bt_h)$ the
value $\D{\f{\beta_g}{2}}=0.250,$ which characterizes the $U(1)$ symmetry.
Thus the system moves from $S$ to $H_4.$
An intriguing feature is that the value
$\bt_g=0.5$ is considered as
strong coupling in the $S$ phase and the $H_4$ phase (this paragraph)
and as weak
coupling in the $H_5$ phase (figure 4(a)). This means that it is not just
the  value of a specific coupling that counts, but also the values of
the remaining couplings; in particular the decisive element is the
phase in which the system lies.

Finally figure 4(c) contains data for $\bt^\pr_g=3.5$ and running $\bt_h.$
The space-like plaquette has the value $\D{\f{\beta_g}{4}}=0.125$ for small
$\bt_h,$ so it exhibits $SU(2)$ confinement, consistent with
an $SU(2)$ strong $(S)$ phase.
For large $\bt_h$ it tends to values consistent with $1-\D{\f{1}{5 \bt_g}}=0.60.$
On the other hand $P_T$ starts from strong coupling values about 0.64.
We again remark that
the apparently weak coupling $\beta_g^\prime=3.5$ yields strong coupling
behaviour, since the system is in an appropriate phase, due to the
values of the remaining couplings.
For large $\bt_h,$ $P_T$ tends to the limit $1-\D{\f{1}{5 \bt_g^\pr}}=0.94$
and it appears safe to conclude that the system moves from $S$ to $H_5.$
The essential difference between the two $S-H$ transitions
lies in the fact that $P_T$ does not change at all during the
$S-H_4$ transition, but changes from strong $SU(2)$ to weak $U(1)$
during the $S-H_5$ transition. The link $L_T$ follows the same scheme.

Thus we have explained how the phase diagram has been derived.
We should note the absence of a $SU(2)$ Coulomb phase for
large values of $\beta_{g}^{\prime}.$ It has not been possible to
find it, although we have searched for it up to $\beta_{g}^{\prime} \sim 10$.
This result may be
attributed to the small chosen value of $\beta_{g}.$
The argument goes as follows: let us fix the gauge by
requiring that $U_5(x)=1.$ Then the transverse-like plaquettes
are in principle driven by $\bt_g^\pr.$ If $\bt_g$ is large
enough, it will couple the neighbouring transverse-like hyperplanes,
e.g. the $\hat{5} \hat{1}$ and $\hat{5} \hat{2}.$ However, if
$\bt_g$ is small (which is the case here) these hyperplanes will
decouple and the model will reduce to several copies of the same
two-dimensional $SU(2)$ gauge model, which does not exhibit phase
transitions.

\vspace{0.5cm}

$\bullet$ \hspace{0.1cm}
$\mathbf{ \beta_{g}=1.2}$\vspace*{0.2cm}

\begin{figure}[!h]
\begin{center}
\includegraphics[scale=0.40]{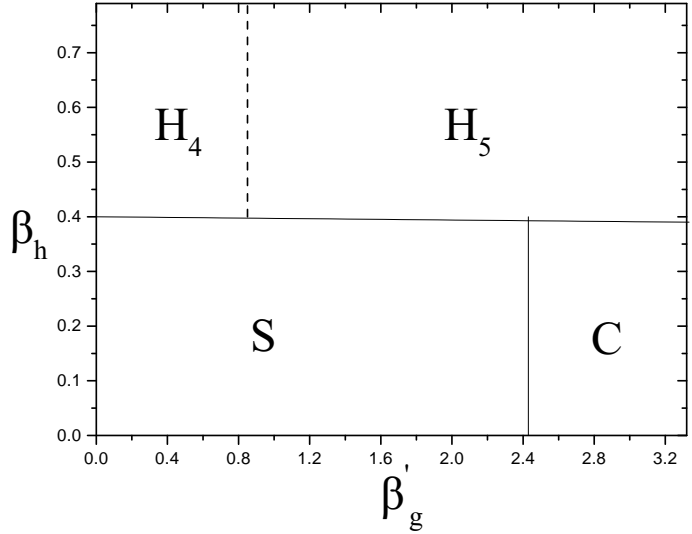}
\caption{Phase diagram for the anisotropic model for $\beta_{g}=1.2$
and for $\beta_{R}=0.01.$
}\label{Fig1.3}
\end{center}
\end{figure}

As we already stated, we also simulated the system at a larger value
of $\bt_g,$ namely $1.2,$ to study its behaviour at weaker couplings.
Two striking new features arise: an $SU(2)$ Coulomb phase and a genuine
layered $(H_4)$ phase with Coulomb (rather than strong) $U(1)$
interations within the layer. As in the previous paragraph, we give the
resulting phase diagram in figure \ref{Fig1.3} and summarize
the behaviours of the plaquettes in the various phases in table 2. Next we will
elaborate on the phase transition lines in more detail and
explain how our claims in figure \ref{Fig1.3} and table 2 are derived.

\begin{table}[!b]
\begin{center}
\hspace*{1cm} \begin{tabular}{|c|c|c|} \hline & \bf{Layer} &
\bf{Transverse direction} \\ \hline \bf{S}& SU(2)--strong: $P_{S}
\sim \D{\f{\beta_{g}}{4}}$ & SU(2)--strong: $P_{T} \sim
\D{\f{\beta^{\prime}_{g}}{4}}$ \\ \hline \bf{C}& SU(2)--Coulomb:
$P_{S} \sim 1-\D{\f{3}{5 \beta_{g}}}$ & SU(2)--Coulomb: $P_{T} \sim
1-\D{\f{3}{5 \beta^{\prime}_{g}}}$ \\ \hline $\mathbf{H_{4}}$ &
U(1)--4D Coulomb: $P_{S} \sim 1-\D{\f{1}{4 \beta_{g}}}$ &
SU(2)--strong: $P_{T} \sim \D{\f{\beta^{\prime}_{g}}{4}}$ \\ \hline
$\mathbf{H_{5}}$ & U(1)--5D Coulomb: $P_{S} \sim 1-\D{\f{1}{5
\beta_{g}}}$ & U(1)--5D Coulomb: $P_{T} \sim 1-\D{\f{1}{5
\beta^{\prime}_{g}}}$  \\ \hline
\end{tabular}
\caption{Table 2: Phase characterization in terms of the $P_{S}$ and $P_{T}$
values for $\beta_{g}=1.2$.}
\end{center}
\end{table}

A new phase transition line is the one separating the strong $S$
phase from the Coulomb $C$ phase. An example of the transition is
given in figure 6(a), where $\bt_h=0.01$ and $\bt_g^\pr$ runs. In
the strong phase $S$ both $P_S$ and $P_T$ take a value equal to
the $\D{\f{1}{4}}$ of their respective lattice couplings, while their
values get large when the system is passes over to the Coulomb
phase $C.$ In particular, $P_S$ tends to the value $1-\D{\f{3}{5
\beta_{g}}}=0.5,$ while $P_T$ follows the curve $1-\D{\f{3}{5
\beta_{g}^\pr}}.$ The
$S-C$ phase transition appears to be first order since there is a
clear hysteresis loop for the relevant points although not a
very large one.

In figure 6(b) we show $P_S, P_T$ and $L_T$ as the system moves
from $H_4$ to $H_5.$ There is no significant hysteresis loop in any of the
three quantities, so the transition is presumably a continuous
phase transition or a crossover. The transverse-like link $L_T$
moves from a small value in $H_4$ to a large value in $H_5,$ which
is actually what one would expect. One can make more quantitative
remarks about the two plaquettes. The space-like plaquette $P_S$
does not really feel the phase transition very much. The only
aspect that changes, from the point of view of $P_S,$ is that the
four-dimensional space-time becomes five-dimensional. On the basis
of what has been already mentioned, in $H_4$  the space--like
plaquette, $P_S$, should approach the
value $1-\D{\f{1}{4 \beta_{g}}}=0.79,$ while in $H_5$ the value $1-\D{\f{1}{5
\beta_{g}}}=0.83.$ Thus one would expect a slight increase in the
value of $P_S$ during the phase transition which is actually observed.
On the other
hand, the transverse-like plaquette $P_T$ starts with the strong
coupling value $\D{\f{\bt_g^\pr}{4}},$ which is expected for the
confining $SU(2),$ and ends up with $1-\D{\f{1}{5 \beta_{g}^\pr}},$
which characterizes the five-dimensional Coulomb $U(1)$ associated
with the $H_5$ phase.

As we already said, a very important point is the emergence of a
genuine layered phase. The transition $S-H_4$ is shown in figure
6(c) and illuminates some properties of $H_4.$ To begin with,
$R^2$ starts from a small value and ends up with values much
larger than 1. This ensures that the system moves into a Higgs
phase. The space-like plaquette $P_S$ starts with the strong
coupling value $\D{\f{\bt_g}{4}}=0.30$ and tends after the transition
to the four-dimensional Coulomb $U(1)$ value $1-\D{\f{1}{4
\beta_{g}}}
= 0.79.$ This is in sharp contrast with the behaviour of $P_S$ in
$H_4$ in the $\bt_g=0.5$ case, figure 4(b), where it followed $\D{\f{\bt_g}{2}},$
implying strong $U(1).$ Here we observe
Coulomb behaviour within the layers.
In addition, we have used the value $D=4$ for
the dimension of space-time to get the value 0.79. If we had used
the value $D=5,$ the resulting value would be 0.83. Our results
are not conclusive in this respect. The Monte Carlo value is
smaller than the analytical prediction anyway.
However comparison with figure 6(b) shows that $P_{S}$ is
relatively small, since in the $H_{5}$ phase its value would
approach 0.8. Thus, it seems that the choice D=4 is correct.

On the other hand one may be
confident that the layers are essentially four-dimensional
entities, since the values of the transverse-like observables are very
small and suggest that the layers are decoupled. In particular
$P_T,$ shown in the same figure, takes the value
$\D{\f{\beta^{\prime}_{g}}{4}},$ signaling $SU(2)$ confinement in the
transverse direction. Thus we have a layered phase, where
particles and gauge fields interact through Coulomb $U(1)$
interactions within the four-dimensional layers and cannot escape
towards the transverse direction because of the $SU(2)$
confinement forces which prevail in the space between the layers.
In figure 6(d) we show the space-like $(L_S)$ and timelike $(L_T)$
links respectively. It is clear that $L_{T}$ has a small value
throughout, while $L_{S}$ grows. This is because the real phase
change takes place within the layer, where the strong $SU(2)$
transforms to four-dimensional $U(1)$ and space-like
quantities, such as $L_S,$ are sensitive to this transition.
The existence of the large hysteresis loops suggests a
first order phase transition.

Some sample results concerning the transition $S-H_5$ are
contained in figures 7(a) and 7(b). In figure 7(a) the quantity
$R^2$ shows that the system moves to a Higgs phase. The space-like
plaquette starts with $SU(2)$ confinement behaviour and ends up
with Coulomb $U(1)$ behaviour.
Similarly, $P_T$ starts with $SU(2)$ confinement
behaviour and ends up with Coulomb $U(1)$ behaviour.
This is consistent with an $S-H_5$ transition. The
analytical predictions here are not precise at all in this case.
In particular, the predictions for the plaquettes in the $H_5$ phase
are $P_S \approx 1-\D{\f{1}{5 \beta_{g}}}=0.83$ and $P_T \approx
1-\D{\f{1}{5 \beta_{g}^\pr}}=0.87.$ However the Monte Carlo results
show that $P_S > P_T.$ Also the analytical result
$\D{\f{\bt_g^\pr}{4}}=0.375$ in the phase $S$ does not agree well with
the Monte Carlo result 0.48. Thus in this case one should be
content with the qualitative features characterizing the phase
transition. The link variables of figure 7(b) are also consistent
with an $S-H_5$ transition. It is remarkable that although we
insist on the rather small value $\beta_{h}^{\prime}=0.001,$ the system
passes over to the $H_{5}$ phase which is characterised by large
values for all order parameters. In addition, the large
hysteresis loops suggest a first order transition.

Finally the $C-H_5$ transition, which is shown in figure 7(c) also
appears to be first order. However the hysteresis loop is small, so
more detailed analysis is needed in this case.

\begin{figure}[!h]
\subfigure[]{\includegraphics[scale=0.25]{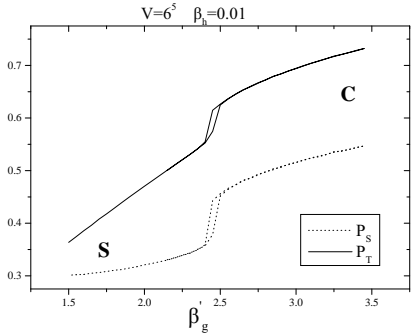}}
\subfigure[]{\includegraphics[scale=0.25]{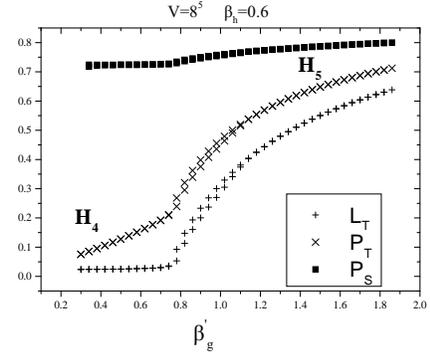}}
\subfigure[]{\includegraphics[scale=0.25]{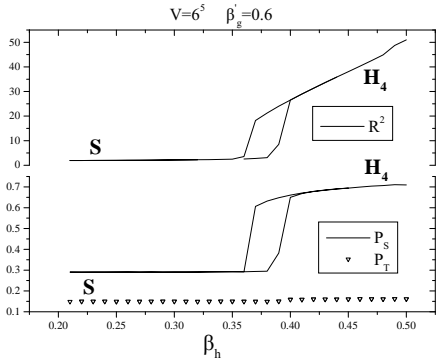}}
\hspace*{3.5cm} \subfigure[]{\includegraphics[scale=0.25]{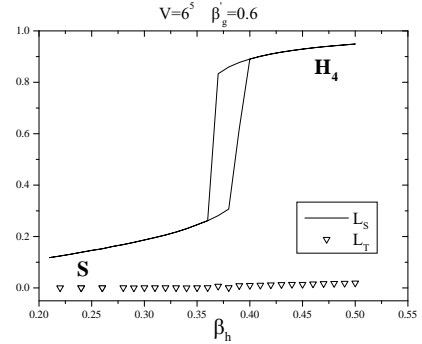}}
\caption{(a) $P_{S}$ and $P_{T}$ for the $S-C$ phase transition.
(b) $L_{T}$, $P_{S}$ and $P_{T}$ for the $H_{4}-H_{5}$
phase transition. (c) $R^2, P_{S}$ and $P_{T}$ for the $S-H_{4}$
phase transition. (d) $L_{S}$ and $L_{T}$ for the $S-H_{4}$
phase transition.}
\label{Fig3}
\end{figure}

\begin{figure}
\subfigure[]{\includegraphics[scale=0.25]{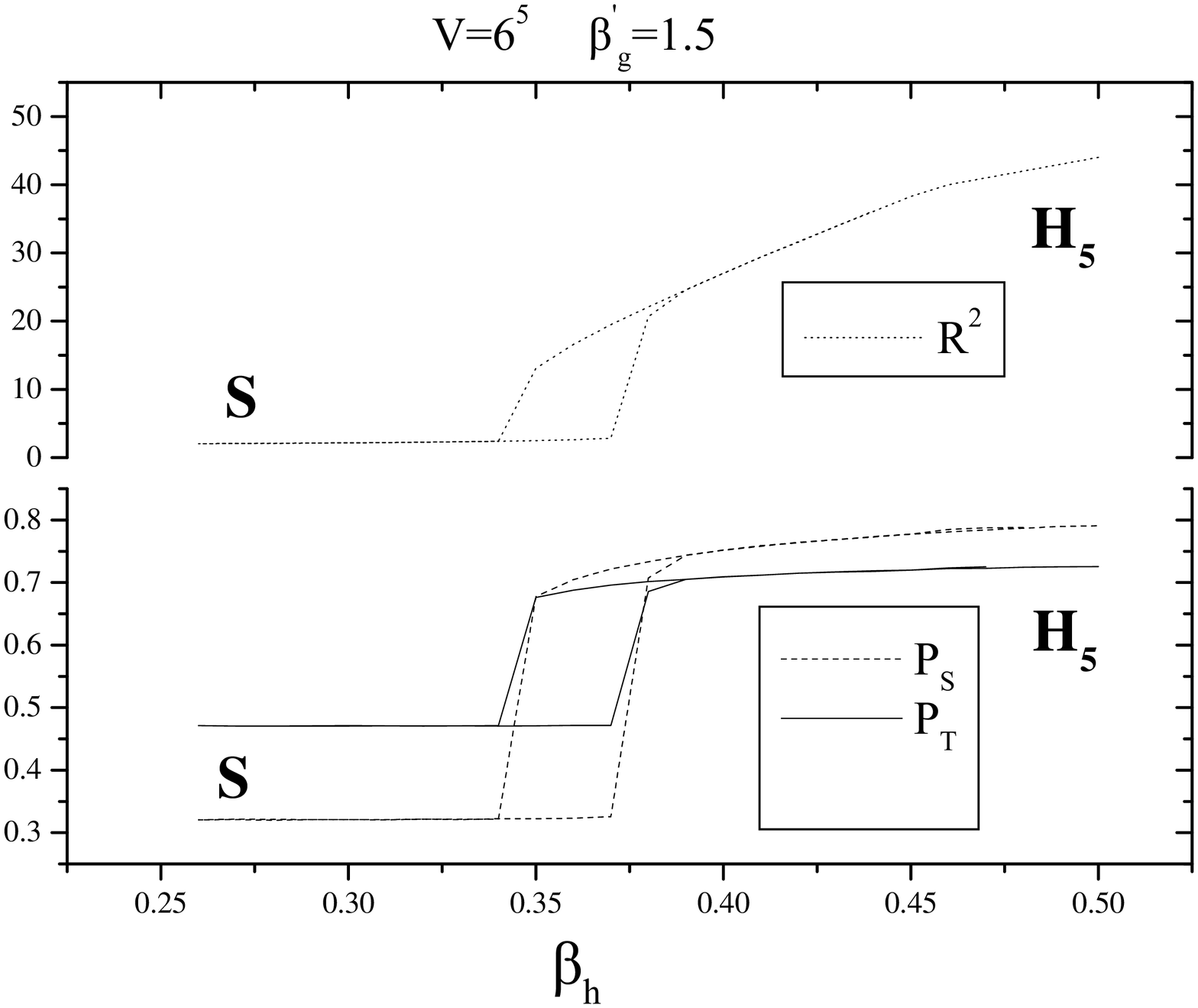}}
\subfigure[]{\includegraphics[scale=0.25]{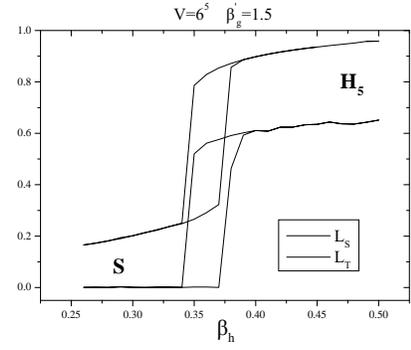}}
\subfigure[]{\includegraphics[scale=0.25]{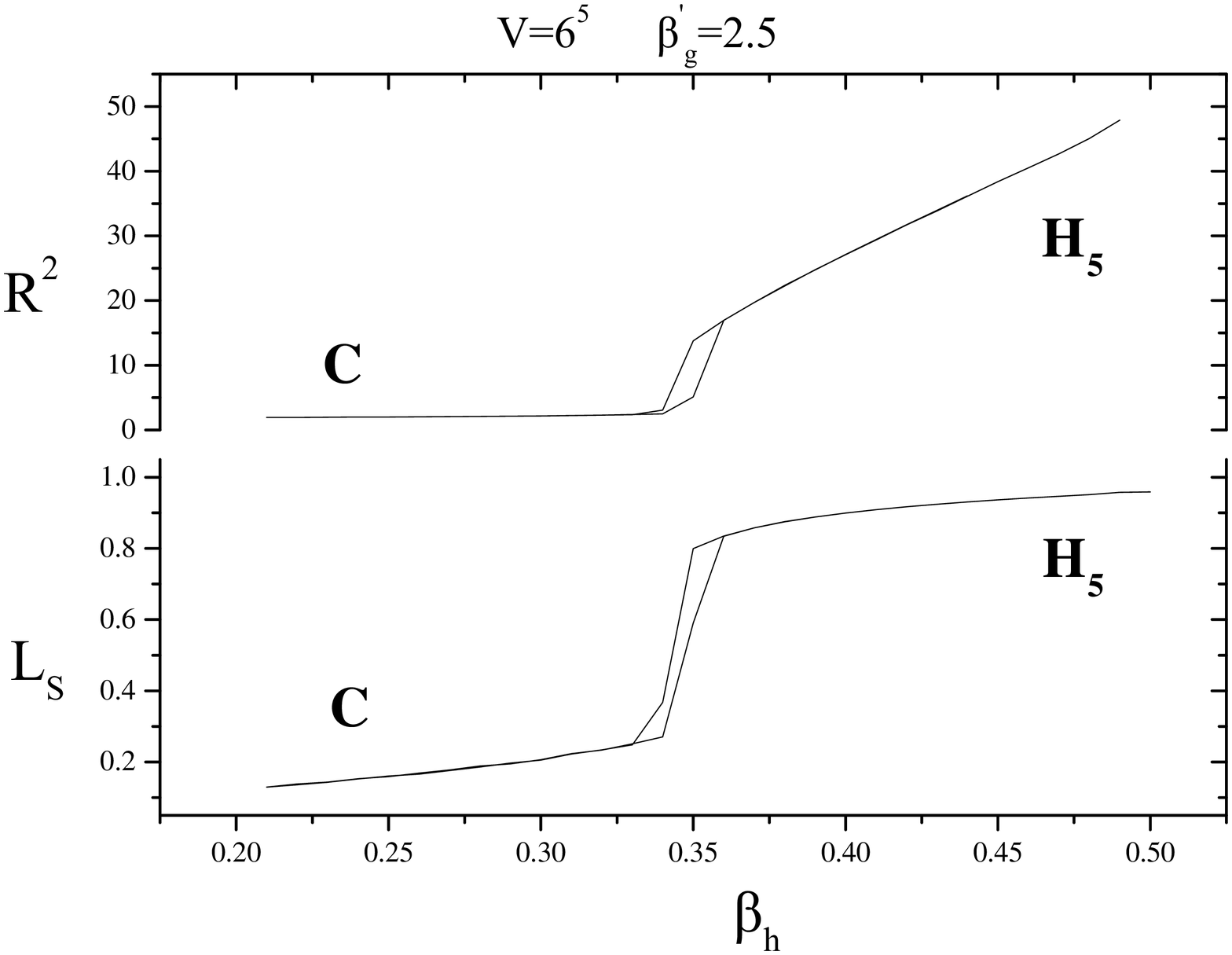}} \caption{(a)
$R^2, P_{S}$ and $P_{T}$ for the $S-H_{5}$ phase transition.  (b)
$L_{S}$ and $L_{T}$ for the $S-H_{5}$ phase transition. (c) $R^2$
and $L_{S}$ for the $C-H_{5}$ phase transition.}
\end{figure}

\subsection{The effect of $\bt_h^\pr$}

\begin{figure}[!h]
\begin{center}
\includegraphics[scale=0.40]{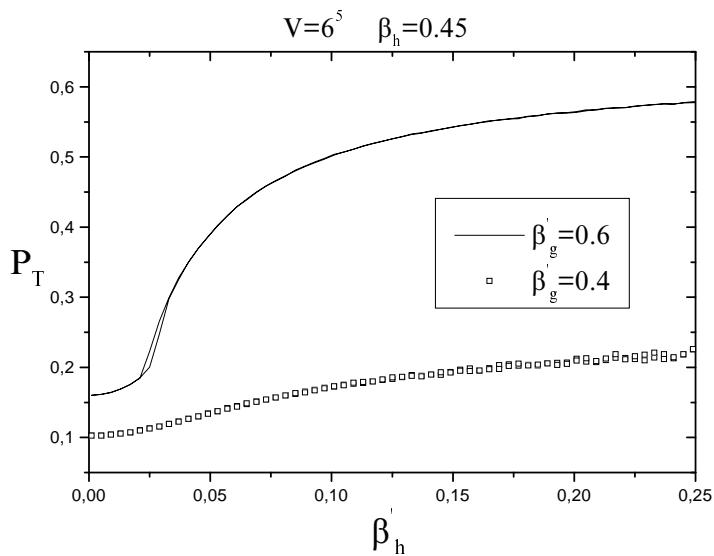}
\caption{The transverse-like plaquette at $\bt_g=1.2, \bt_h=0.45$ and $\bt_R=0.01$ for $\bt_g^\pr=0.4$ and $\bt_g^\pr=0.6.$ }
\label{ptbgt0406bhtrun}
\end{center}
\end{figure}

\begin{figure}[!h]
\begin{center}
\includegraphics[scale=0.40]{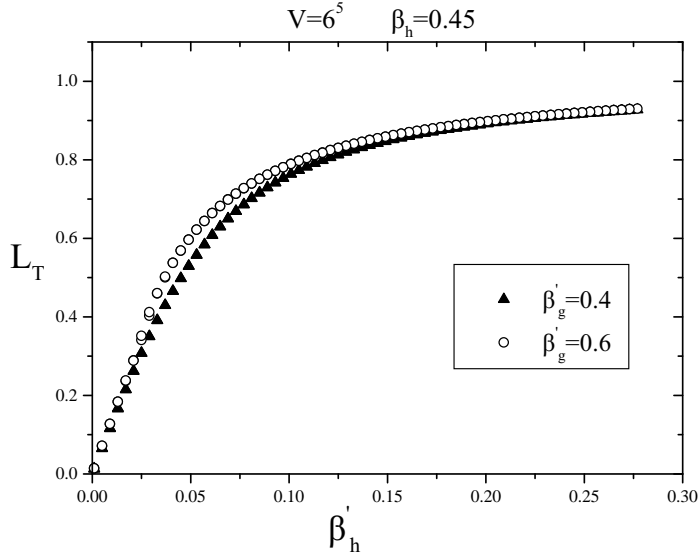}
\caption{The transverse-like link at $\bt_g=1.2, \bt_h=0.45$
and $\bt_R=0.01$ for $\bt_g^\pr=0.4$ and $\bt_g^\pr=0.6.$ }
\label{ltbgt0406bhtrun}
\end{center}
\end{figure}

In this paragraph we go ahead with some results concerning the
effects of the $\bt_h^\pr$ coupling, which has been set to the
value 0.001 up to now. In figure \ref{ptbgt0406bhtrun} we have
fixed the couplings to the values $\bt_g=1.2, \ \bt_h=0.45, \
\bt_R=0.01$ and performed a scanning in $\bt_h^\pr$ at two values
of $\bt_g^\pr.$ The order parameter considered is the
transverse-like plaquette $P_T.$ For $\bt_g^\pr=0.4$ we see that
we start from an $SU(2)$ confining theory for small $\bt_h^\pr$
(the value of the plaquette equals $\D{\f{\bt_g^\pr}{4}}=0.1)$ to end
up with a $U(1)$ confining theory, since the value of $P_T$ tends
to $\D{\f{\bt_g^\pr}{2}}=0.2.$
The space--like plaquette $P_{S}$ takes on big values and it seems
to indicate that the layers remain in the Coulomb phase.
One may be tempted to think that it is
a layered phase. However, a look at figure \ref{ltbgt0406bhtrun},
which contains the corresponding links, reveals something strange:
for large enough $\bt_h^\pr$ the transverse-like link is not
small, as one would expect from a layered phase. The physical
understanding of this situation is that the gauge field cannot
really travel from a layer to the neighbouring one. However, since
we have arranged the couplings to place the system in the Higgs
phase, symmetry breaking has occured, resulting in a $U(1)$
residual symmetry along with a scalar particle with zero charge.
Nothing can prevent this particle from moving over the whole
lattice, which has the result of giving a large value for $L_T,$
as we see in figure \ref{ltbgt0406bhtrun}. We remark that $L_T$ is
the gauge invariant propagator of the scalar particle at the
distance of one lattice spacing. This physical situation has not
been encountered before. It is a new phase, which contains $U(1)$
gauge fields localized within the layers and a freely moving
scalar field. One may like to call this phase $C_4,$ since the
scalar field does not play any significant role. It is a gauge field
in a four-dimensional Coulomb phase coupled to a massive scalar.

At $\bt_g^\pr=0.6$ we have the upper curve of figure
\ref{ptbgt0406bhtrun}. The system starts at the $SU(2)$ confining
phase with $P_T$ taking the value $\D{\f{\bt_g^\pr}{4}}=0.15,$ but the
final phase appears to be weakly coupled $U(1),$ with $P_T \approx
1-\D{\f{1}{5 \bt_g^\pr}} = 0.67.$ The transverse-like link is also
large in this case for the same reason. If one would like to give
a name to this phase, one would call it $C_5.$ We observe that,
although at $\bt_g^\pr=0.4$ there is no trace of a phase
transition between strong $SU(2)$ and strong $U(1),$ the
transition from strong $SU(2)$ to weak $U(1)$ at $\bt_g^\pr=0.6$
exhibits a (small) hysteresis loop in between, signaling a
possible phase change. We remark here that the value
$\bt_h^\pr=0.001$ that we used for most of our Monte Carlo
simulations is rather small; in particular it lies before the
phase transition for $\bt_g^\pr=0.6$ of figure
\ref{ptbgt0406bhtrun}.

A remark concerning figure \ref{ltbgt0406bhtrun} is that the curve for for
$\bt_g^\pr=0.6$ is steeper than the one for $\bt_g^\pr=0.4.$ Thus
we expect (actually verified) that for bigger values of $\bt_g^\pr$
it will be even steeper, meaning that if $\bt_g^\pr$ is large enough,
even a small value of $\bt_h^\pr$ is enough to take the system out of the
$SU(2)$ confinement.

\begin{figure}[!h]
\begin{center}
\includegraphics[scale=0.40]{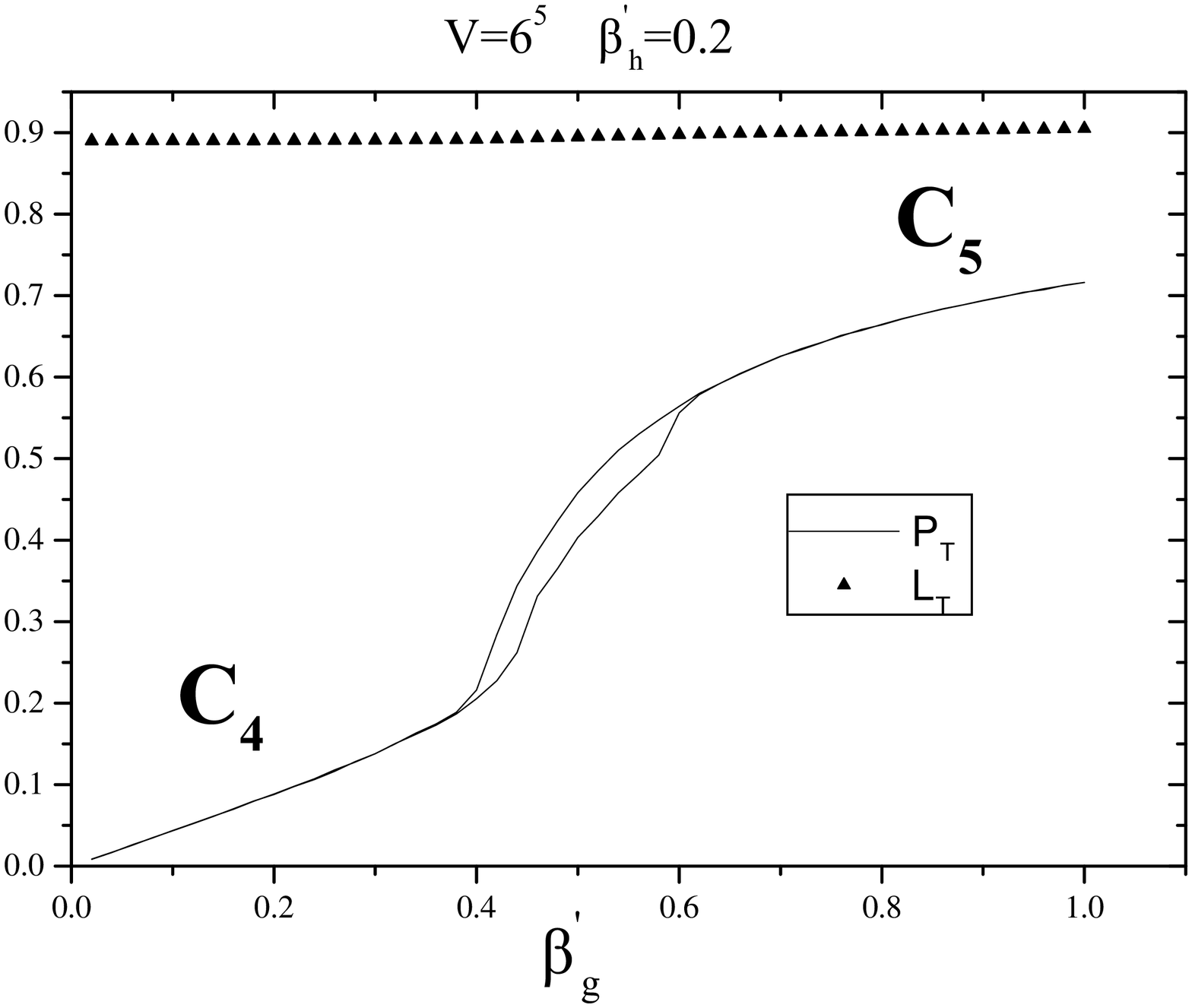}
\caption{The transverse-like plaquette and link at $\bt_g=1.2, \bt_h=0.45$
and $\bt_R=0.01$ for $\bt_h^\pr=0.2$ and running $\bt_g^\pr.$ }
\label{ptltbht02bgtrun}
\end{center}
\end{figure}

From figure \ref{ptbgt0406bhtrun} we see that if we set
$\bt_h^\pr=0.20$ and let $\beta_g^\pr$ run, we should see a change
in the behaviour of the system between $\bt_g^\pr=0.4$ and
$\bt_g^\pr=0.6.$ The results of this run is displayed in figure
\ref{ptltbht02bgtrun}. The transverse-like link takes a large
value, the same in both the $C_4$ and $C_5$ phases, while $P_T$
follows $\D{\f{\bt_g^\pr}{2}}$ in the beginning and then changes to
$1-\D{\f{1}{5 \bt_g^\pr}},$ with a hysteresis loop in between.

\begin{figure}
\subfigure[]{\includegraphics[scale=0.25]{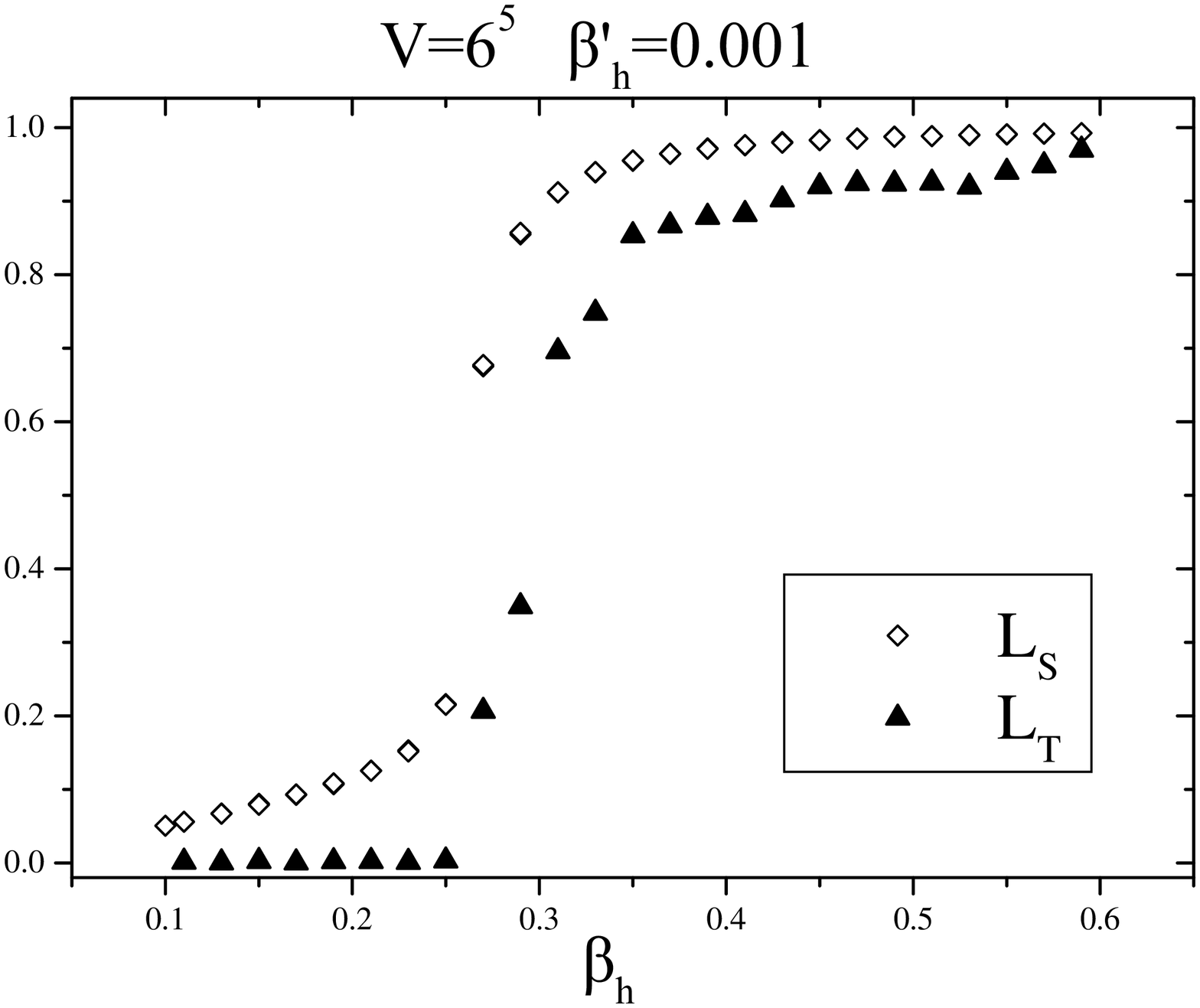}}
\subfigure[]{\includegraphics[scale=0.25]{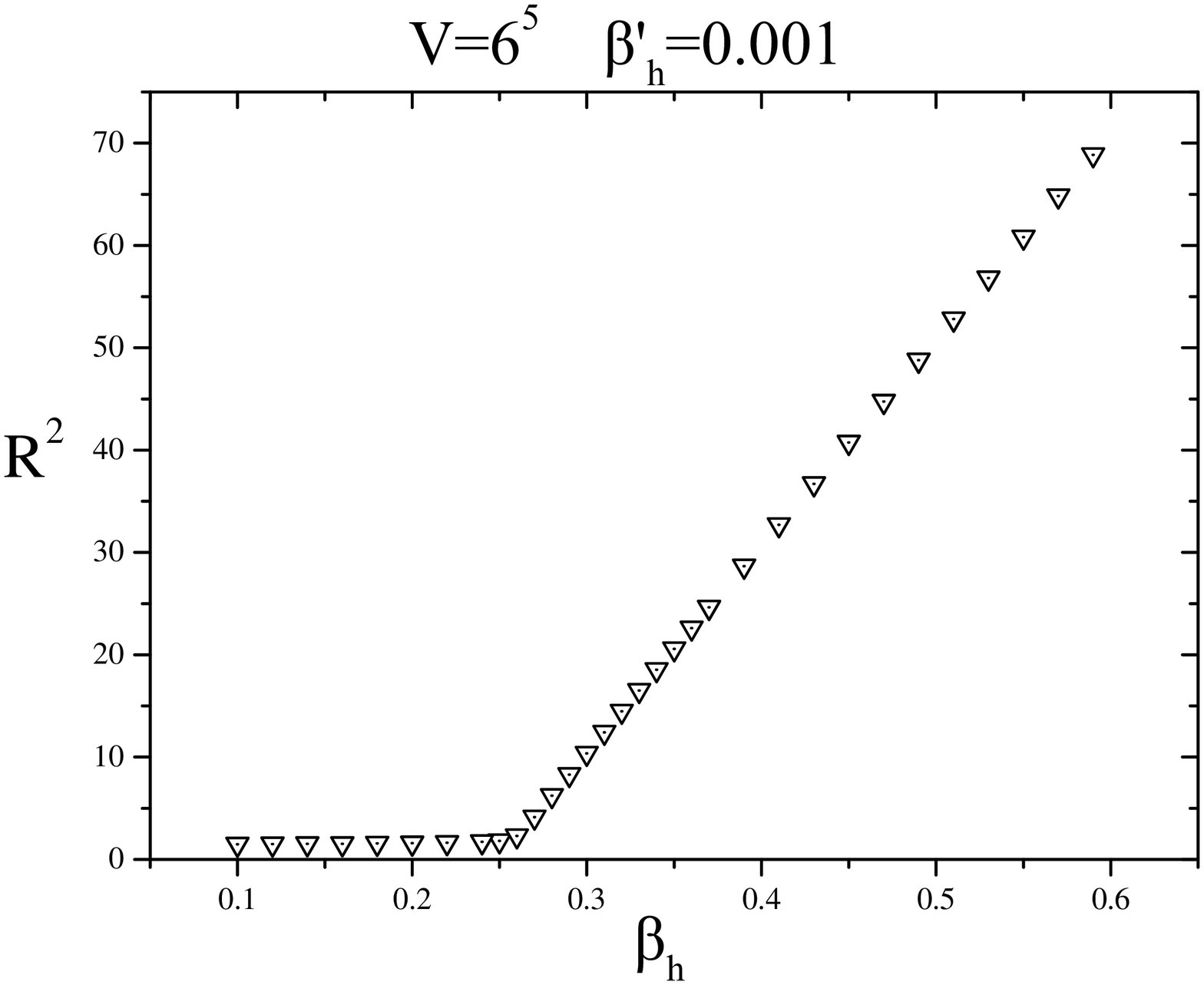}}
\caption{Graphs derived by hysteresis loop technique for the pure
scalar field showing the $L_{S}$, $L_{T}$ (fig.(a)) and $R^{2}$
(fig. (b)) as $\beta^{\prime}_{h}=0.001$ and $\beta_{h}$ is
running.} \label{scalar}
\end{figure}

\subsection{The role of the scalar field}

We would like to study the role of the scalar field in the
formation of the layered phase, so we simulate our model setting
the lattice gauge couplings $\bt_g$ and $\bt_g^\pr$ to infinity.
Thus the gauge fields do not play
role any more and we are left with the dynamics of the scalar
fields. In a sense figure \ref{scalar}, which we will present
shortly, is similar to figure 7(c), with the difference that
$\bt_g$ and $\bt_g^\pr$ are infinite rather than taking a finite
value. The results on the hysteresis loops for $L_{S}$ and $L_{T}$
of the pure scalar field model are shown in figure \ref{scalar}.
We have set $\beta_{h}^{\prime}=0.001,$ while $\beta_{h}$ is
running. One can see that there is a possible phase transition
near $\beta_{h}=0.25$ (which is a reasonable critical point, since
one would expect the value $\D{\f{1}{D-1}},$ with $D=5$). Before the
phase transition we have a symmetric phase (since $R^2$ is
small, as shown in figure \ref{scalar}(b)); in addition $L_T$ is
zero in this phase, so we conclude that the system actually
consists of four-dimensional layers.
After the transition $R^2$ becomes large and the two links take
large, almost equal values. Thus, the system moves presumably towards a
five-dimensional Higgs phase, $H_5.$

An interesting issue is the dependence of $L_T$ on the value of
$\bt_h^\pr.$ In figure \ref{scalar}(a) (that is, for $\bt_h^\pr=0.001)$
the transverse-like link $L_T$ takes
a large value if $\bt_h$ becomes
large enough. Thus it appears that no $H_4$ phase appears. On the other hand
we know that if $\bt_h^\pr=0$ the scalar model is strictly four-dimensional.
Thus we expect some phase transition between the two values
$\bt_h^\pr=0$ and $\bt_h^\pr=0.001.$
The question is whether the phase transition takes place exactly at
$\bt_h^\pr=0$ or at some small but non-zero value.
To this end we have measured $L_T$
for small values of $\bt_h^\pr$ and present the results in table 3.
For each value of $\bt_h^\pr$ we present the statistical average of $L_T$
in the second column along with the limits within which this quantity
fluctuates in columns three and four.
One may easily see that for $\bt_h^\pr$ smaller than about
$10^{-5}$ the transverse-like link $L_T$ is compatible with zero,
while the fluctuations are large.
The mean value $L_T$ increases with $\bt_h^\pr$, while the fluctuation
range decreases.

It appears that $L_T$ is zero for small $\bt_h^\pr.$ It seems that
there exists a critical value for $\bt_h^\pr$ above which $L_T$ takes
on a non-zero value. However, it is possible that the existence of a
non-zero critical value might be a volume effect. In particular this
critical value (which lies between $10^{-5}$ and $10^{-4}$ for the
$6^5$ volume considered in table 4) may tend to zero as the volume is
increased. In other words it is possible to have this phase change
exactly at $\bt_h^\pr=0.$ To investigate this issue we have set
$\bt_h^\pr$ to two small values, namely $10^{-4}$ and $10^{-5}$ and
measured $L_T$ for various volumes. For $\bt_h^\pr=10^{-4}$ we see in
table 4 that we have wild fluctuations and a small mean value for the
$4^5$ volume, while for bigger volumes the fluctuations are drastically
reduced and the average increases accordingly. Thus, if one would stop
at $4^5,$ one might think that the coupling $\bt_h^\pr=10^{-4}$
lies in the region where $L_T=0.$ However, we see readily that this
depends very much on the volume used: for larger volumes the same
coupling lies in the regime where $L_T \ne 0.$
One may try to move to even smaller values for $\bt_h^\pr,$ for example
the value $10^{-5}$ shown on the right part of table 4.
This value appears to be in the $L_T=0$ region for both $4^5$ (not shown)
and $6^5$ lattices. However, if we move to the $8^5$ volume, it also
gives non-zero values for $L_T.$ It seems safe to conclude that the
critical value of $\bt_h^\pr$ for infinite volume is actually $\bt_h^\pr=0.$
Perhaps one should also explore smaller values of $\bt_h^\pr$ to
corroborate this conclusion, however we claim that, since we have
checked down to values of the order of $\D{\f{1}{V}},$ this conclusion holds
true.

\begin{table}[!h]
\begin{center}
\hspace*{1cm} \begin{tabular}{|c|r|r|r|} \hline
$\bt_h^\pr$ & $L_T$ & $L_{low}$ & $L_{high}$\\ \hline
$0.00$    & -0.059& -0.65 &0.46\\ \hline
$10^{-7}$& -0.075& -0.69 &0.44\\ \hline
$10^{-5}$&  0.015& -0.60 &0.75\\ \hline
$10^{-4}$&  0.745&  0.41 &0.93\\ \hline
$10^{-3}$&  0.948&  0.89 &0.98\\ \hline
$10^{-2}$&  0.968&  0.91 &0.99\\ \hline
\end{tabular}
\end{center}
\caption{Behaviour of $L_T$ and the limits of its statistical
fluctuations for the pure scalar field for a $6^5$ lattice.}
\end{table}

\begin{center}
\begin{table}[!h]
\begin{center}
\hspace*{1cm} \begin{tabular}{|c|r|r|r||c|r|r|r|} \hline
\multicolumn{4}{|c||}{$\bt_h^\pr = 10^{-4}$} & \multicolumn{4}{c|}{$\bt_h^\pr = 10^{-5}$} \\ \hline
V& $L_T$ & $L_{low}$ &$L_{high}$ & V& $L_T$ & $L_{low}$ & $L_{high}$ \\ \hline
$4^5$ & 0.258 & -0.54 & 0.94 & $6^5$ & 0.15 & -0.60 & 0.75 \\ \hline
$6^5$ & 0.745 &  0.41 & 0.93 & $8^5$ & 0.890 & 0.80 & 0.95 \\ \hline
$8^5$ & 0.934 &  0.87 & 0.96 & $10^5$& 0.915 & 0.83 & 0.97 \\ \hline
\end{tabular}
\end{center}
\caption{Behaviour of $L_T$ and the limits of its statistical
fluctuations for the pure scalar field. Two values for $\bt_h^\pr$ are
explored for various lattice volumes.}
\end{table}
\end{center}

In conclusion, we have not found any sign of four-dimensional
(layered) Higgs phase for the scalar model. Its absence
can be attributed to the lack of gauge interactions
which could provide the mechanism for confinement along the
transverse direction.
This leads to the conclusion that
if one wants to have a layered phase in realistic theories with
localization of the fields within a
four--dimensional subspace (in the framework of a theory in higher
dimensions), gauge field interactions are necessary.
Similar conclusions have been reached by Fu and Nielsen \cite{fn2} through a
mean field treatment of the $O(N)$ model. The ``localization scale"
(i.e. the energy above from which the extra
dimension becomes visible) appears to be determined by the gauge
coupling constants. On the other hand, we have already mentioned that
a pure $SU(2)$ gauge theory has no layer phase in five dimensions. It appears
that a coordinated action of the gauge and scalar sectors is
necessary to produce a layered phase in 5 dimensions, since each sector
on its own cannot be efficient in this respect.

\newpage
\section{Concluding Remarks}
In this paper we have tried to find a layer phase in the SU(2)--adjoint
Higgs model in five dimensions using lattice simulations. Apart
from exploring the phase diagram for the isotropic model we
focused our attention to the {\it anisotropic} one. We determined
the phase diagram in the $\beta_{h} - \beta_{g}^{\prime}$
hyperplane of the lattice parameter space and we found out
that there is a four--dimensional layered Higgs phase exhibited by
the model which is separated by a phase boundary from the
strong phase. This is an entirely new result
which is attributed to the gauge--scalar interaction considered.
Let us recall the well-known result that
the pure SU(2) gauge theory exhibits a layered structure
in six dimensions only and yields a non-physical Coulomb
layered phase in five dimensions. The other new and interesting
feature of the model is that for some region of the lattice
couplings the formation of the layered Higgs phase is
attributed to the SU(2) strong interaction along the
transverse--extra dimension. This fact serves as an important
physical indication of the way that four dimensional layers
may be formed. \\
The whole construction may have interesting implications on models
based on grand unified groups, such as $SU(5),$ with the scalars in
the adjoint representation (for example the one with dimension 24)
defined in a  higher--dimensional space.  From this model
we could get a layered phase in four dimensions. The model would be
promising if the relevant scale of this layer formation would be
of the same order of magnitude as the scale of
the symmetry breaking from $\mbox{SU(5)}$ down to
$\mbox{SU(3)} \times \mbox{SU(2)} \times \mbox{U(1)}$.

\vspace*{0.7cm}
\noindent{\Large \bf{Acknowledgements}}

\noindent The authors acknowledge partial financial support by the TMR
Network entitled ``Finite Temperature phase transitions in
Particle Physics", EU contract number FMRX-CT97-0122. K.F. thanks
C. Bachas and C. Korthals--Altes for useful discussions.

\vspace*{1.2cm}

\noindent{\Large \bf{Appendix: Mean Field Analysis}}
\vspace*{0.3cm}

\noindent Our starting point is the partition function \be Z=\prod_x
\int \prod_{\mu=1}^5 dU_\mu(x) \int d \Phi(x)
e^{-S(a_\mu^M, e^m)}, \quad S(a_\mu^M, e^m) = S_G+S_{GH}+S_H, \label{pf} \ee
where $$S_G = -\f{\bt_g}{2} \sum_{x,
\mu<\nu<5} Tr [U_\mu(x) U_\nu(x+\hat{\mu}) U^\dagger_\mu(x+\hat{\nu})]
U^\dagger_\nu(x) -\f{\bt_g^\pr}{2} \sum_{x, \mu<5} Tr [U_\mu(x)
U_5(x+\hat{\mu}) U^\dagger_\mu(x+\hat{5}) U^\dagger_5(x)],$$ $$S_{GH} =
-\f{\bt_h}{2} \sum_{x,\mu<5} Tr[\Phi(x) U_\mu(x) \Phi(x+\hat{\mu})
U^\dagger_\mu(x)]-\f{\bt_h^\pr}{2} \sum_{x} Tr[\Phi(x) U_5(x)
\Phi(x+\hat{5}) U^\dagger_5(x)],$$ $$S_H = \bt_R \sum_x\left(\f{1}{2}
Tr \Phi^2-1\right)^2.$$ We note in addition that $$\Phi(x) = \phi^m(x)
\sigma^m, \phi^m(x) \equiv \rho(x) e^m(x)
= \rho(x) (\sin \theta(x) \cos \psi(x), \sin
\theta(x) \sin \psi(x), \cos \theta(x))$$
and $$d \Phi(x) = \f{1}{\pi^{3/2}} e^{-\rho^2(x)} \rho^2(x) d
\rho(x) d \cos \theta(x) d \psi(x) \equiv
e^{-\rho^2(x)} \rho^2(x) d \rho(x) d e(x),$$ along with $$U_\mu(x)=
a^0_{\mu}(x)+i \sum_{m=1}^3 a_\mu^m(x) \sigma^m,$$ $$dU_\mu(x) = \f{1}{\pi^2}
\delta[(a^0_{\mu}(x))^2+\sum_{m=1}^3 (a_\mu^m(x))^2-1] \prod_{M=0}^3
d a_\mu^M(x).$$

We introduce the unconstrained variables $A_\mu^M(x), M=0,1,2,3,$
corresponding to the gauge field variables
$a_\mu^M(x), M=0,1,2,3,$ and use the identity: $$1
= \prod_x \prod_{M=0}^3 \prod_{\mu=1}^5 \int_{-\infty}^{+\infty} d A_\mu^M(x)
\delta(A^M_\mu(x)-a^M_\mu(x)) =$$
$$ = \prod_x \prod_{M=0}^3 \prod_{\mu=1}^5 \int_{-\infty}^{+\infty}
d A_\mu^M(x)
\int_{-i \infty}^{+i \infty}
\f{d \alpha^M_\mu(x)}{2 \pi i} \exp[-\alpha^M_\mu(x)
(A^M_\mu(x)-a^M_\mu(x))],$$ We follow a similar path with the
unconstrained variables $E^m(x), m=1,2,3,$ corresponding to
the scalar field variables
$e^m(x), m=1,2,3,$ which satisfy the identity: $$1 =
\prod_x \prod_{m=1}^3 \int_{-\infty}^{+\infty} d E^m(x)
\delta(E^m(x)-e^m(x)) =$$ $$= \prod_x \prod_{m=1}^3 \int_{-\infty}^{+\infty}
d E^m(x) \int_{-i \infty}^{+i \infty} \f{d \epsilon^m(x)}{2 \pi i}
\exp[-\epsilon^m(x) (E^m(x)-e^m(x))].$$

Next we insert the above factors of 1
in the partition function (\ref{pf}) and rearrange the order of the
integrations:

$$
Z =
\int_{-\infty}^{+\infty} \prod_{M=0}^3 \prod_{\mu=1}^5 d A_\mu^M(x)
\int_{-i \infty}^{+i \infty} \f{d \alpha^M_\mu(x)}{2 \pi i}
\int_{-\infty}^{+\infty} \prod_{m=1}^3 d E^m(x)
\int_{-i \infty}^{+i \infty} \f{d \epsilon^m(x)}{2 \pi i}
$$
$$
\exp[-S(A_\mu^M,E^m)-\sum_{M=0}^3 \sum_{\mu=1}^5 \sum_x
\alpha_\mu^M(x) A_\mu^M(x)
-\sum_{m=1}^3 \sum_x \epsilon^m(x) E^m(x)]
$$
$$
\int \prod_x \prod_{\mu=1}^5 dU_\mu(x)
\exp\left[\sum_{M=0}^3 \sum_{\mu=1}^5 \sum_x \alpha_\mu^M(x) a_\mu^M(x)\right]
\int \prod_x d \Phi(x)
\exp\left[\sum_{m=1}^3 \sum_x \epsilon^m(x) e^m(x)\right].
$$
Notice that we have used the delta functions to replace $S(a_\mu^M,e^m)$
by $S(A_\mu^M,E^m).$ The two integrals which appear in the last
line can actually be computed.
Thus
$$\exp[\zeta_G(\alpha_\mu^M(x))] \equiv
\int \prod_x \prod_{\mu=1}^5 dU_\mu(x)
\exp\left[\sum_{M=0}^3 \sum_{\mu=1}^5 \sum_x \alpha_\mu^M(x) a_\mu^M(x)
\right]$$ and
$$\exp[\zeta_H(\epsilon^m(x))] \equiv
\f{1}{4 \pi}
\int \prod_x d e(x) \exp\left[\sum_{m=1}^3
\sum_x \epsilon^m(x) e^m(x)\right]$$
are known functions.

At this stage we make a translationally invariant ansatz, namely
$$A_\mu^M(x)=A \delta_{M0}, \ \alpha_\mu^M(x) = \alpha \delta_{M0}, \ \mu<5,
\ M=0,1,2,3, $$
$$A_5^M(x)=A^\pr \delta_{M0}, \ \alpha_5^M(x) = \alpha^\pr \delta_{M0}, \
M=0,1,2,3, $$
$$\sqrt{[E^1(x)]^2+[E^2(x)]^2+[E^3(x)]^2}=E,
\ \sqrt{[\epsilon^1(x)]^2+[\epsilon^2(x)]^2+[\epsilon^3(x)]^2} = \epsilon, \
\rho(x)=\rho.$$
One can see that, with this choice, the one-site integrals equal:
$$\exp[\zeta_G(\alpha)] = \f{2 I_1(\alpha)}{\alpha}$$ and
$$\exp[\zeta_H(\epsilon)] = \f{\sinh(\epsilon)}{\epsilon}.$$

Thus the evaluation of the partition function in this approximation reduces
to:

$$
Z=
\int_{-\infty}^{+\infty} d A
\int_{-i \infty}^{+i \infty} \f{d \alpha}{2 \pi i}
\int_{-\infty}^{+\infty} d E
\int_{-i \infty}^{+i \infty} \f{d \epsilon}{2 \pi i}
\int_0^{+\infty} d \rho
$$
$$
\exp[-S(A,A^\pr,E) -4 \alpha A - \alpha^\pr A^\pr
- \epsilon e+4 \zeta_G(\alpha)+\zeta_G(\alpha^\pr) +\zeta_H(\epsilon)
+log[\rho^2]]
$$
which equals
$$
Z=
\int_{-\infty}^{+\infty} d A
\int_{-i \infty}^{+i \infty} \f{d \alpha}{2 \pi i}
\int_{-\infty}^{+\infty} d E
\int_{-i \infty}^{+i \infty} \f{d \epsilon}{2 \pi i}
\int_0^{+\infty} d \rho
$$
$$
\exp[-\{-6 \beta_g A^4-4 \beta_g^\pr A^{\pr 4}
-4 \bt_h A^2 \rho^2 E^2 -\bt_h^\pr A^{\pr 2} \rho^2 E^2
+(1-2 \beta_R) \rho^2+\beta_R \rho^4-\log[\rho^2]
$$
$$
+4 \alpha A-4 \zeta_G(\alpha) + \alpha^\pr A^\pr-\zeta_G(\alpha^\pr)
+ \epsilon E -\zeta_H(\epsilon)) \}]
$$
The expression:
\begin{equation}
\label{fren}
\begin{array}{ll}
V=& -6 \beta_g A^4-4 \beta_g^\pr A^{\pr 4} \\
  & -4 \bt_h A \rho^2 E^2-\bt_h^\pr A^\pr \rho^2 E^2\\
  & +(1-2 \beta_R) \rho^2+\beta_R \rho^4-\log[\rho^2] \\
  & +4 \alpha A-4 \zeta_G(\alpha)+\alpha^\pr A^\pr-\zeta_G(\alpha^\pr)
    +\epsilon E -\zeta_H(\epsilon))]
\end{array}
\end{equation}
represents the effective potential.
In five dimensions there are six space-like plaquettes and four
transverse-like ones; this
explains the first line of expression (\ref{fren}).
The second line contains the expressions for the four space-like
links along directions $\hat 1,~\hat 2,~\hat 3,~\hat 4$
and the transverse-like link. The third line
contains the terms that do not refer to directions at all; in particular the
logarithmic last term comes from the measure of the Higgs field. Finally,
the last line contains the contributions of the
integration of the Haar measure:
four $\alpha A- \zeta_G(\alpha)$ terms and
the term with primed quantities from the transverse-like links. A
similar term is connected with the angle of the scalar field.

Our task reduces now to finding the (absolute) minimum of the effective
potential.
Thus, in the saddle point approximation, we have (for a given $\rho$)
to solve the equations:
\begin{center}
\begin{tabular}{cc}
$\D{\f{\p V}{\p \epsilon}} = 0 \ri E=\coth(\epsilon)-\D{\f{1}{\epsilon}},$
& $\D{\f{\p V}{\p E}} = 0 \ri \epsilon=8 \bt_h \rho^2 A^2 E
+2 \bt_h^\pr \rho^2 A^{\pr 2} E,$\\
$\D{\f{\p V}{\p \alpha}} = 0 \ri A=\D{\f{I_2(\alpha)}{I_1(\alpha)}},$
&$\D{\f{\p V}{\p A}} = 0 \ri \alpha = 6 \bt_g A^3 + 2 \bt_h \rho^2 A
E^2.$\\
$\D{\f{\p V}{\p \alpha^\pr}} = 0 \ri
A^\pr=\D{\f{I_2(\alpha^\pr)}{I_1(\alpha^\pr)}},$ &
$\D{\f{\p V}{\p A^\pr}} = 0 \ri
\alpha^\pr = 16 \bt_g^\pr A^{\pr 3} + 2 \bt_h^\pr \rho^2 A^\pr E^2.$
\end{tabular}
\end{center}
(Of course one should also make sure that it is a minimum and not
another kind of extremum.)
The strategy has been to find first the minimum for fixed
values of $\rho$ and then minimize with respect to $\rho.$
In practice we used the minimization facility of Mathematica rather than
solving the simultaneous equations.

\begin{figure}[!h]
\begin{center}
\includegraphics[scale=0.4]{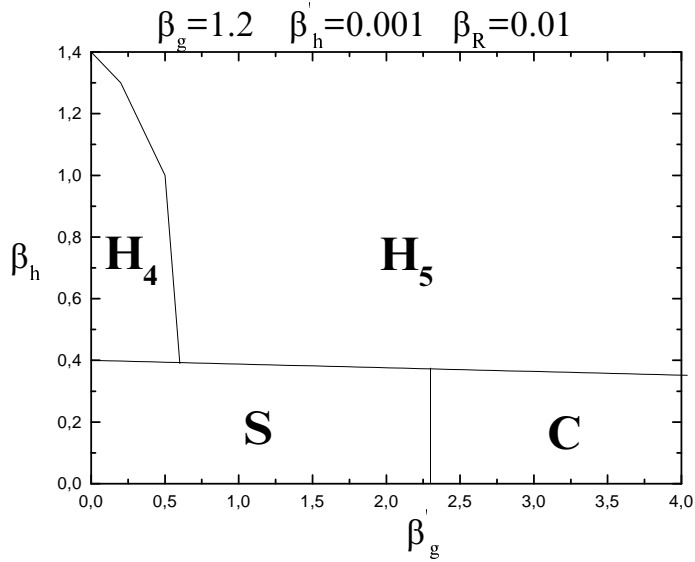}
\caption{Mean field phase diagram in the $\bt_g^\pr - \bt_h$ plane for
$\bt_g=1.2,\beta_{R}=0.01.$ }
\label{mfphd}
\end{center}
\end{figure}

To check whether we find consistent results with the Monte Carlo approach
we now proceed with the study of the anisotropic model setting
$\bt_g=1.2, \bt_h^\pr=0.001, \bt_R=0.01.$ In other words we compare
the Mean Field results with our previous Monte Carlo results. The resulting
phase diagram is shown in figure \ref{mfphd}.

It contains four phases with
the following characteristics:

\begin{center}
\begin{tabular}{llllll}
Quantity:                           &$P_S$   &$P_T$   &$L_S$   &$L_T$   &$R2$ \\
Confining Strong Phase S:           &$0    $ &$0    $ &$0$     &$0   $  &$1$  \\
Coulomb Phase C:                    &$\ne 0$ &$\ne 0$ &$0   $  &$0   $  &$1$  \\
Four-dimensional Higgs Phase $H_4:$ &$\ne 0$ &$0    $ &$\ne 0$ &$0   $  &$>1$ \\
Five-dimensional Higgs Phase $H_5:$ &$\ne 0$ &$\ne 0$ &$\ne 0$ &$\ne 0$ &$> 1$
\end{tabular}
\end{center}

As we vary $\bt_h$ or $\bt_g^\pr$ we see the various quantities changing
to values signaling the transition to some other phase. In general the
phase diagram coincides with the one derived by Monte Carlo runs, apart
from expected quantitative differences. However there are some remarks
to be made. (a) The mean field method always yields a phase transition
separating the $S$ and $C$ phases. The Monte Carlo simulation for
$\bt_g=0.5$ does not exhibit a $C$ phase. It appears that this phase
transition line at strong gauge coupling is present only in
high-dimensional space-times, where the Mean Field approach is more reliable. Of
course, if the gauge coupling is weak, the $C$ phase also exists for relatively
low--dimensional space-times. (b) The phase transitions are predicted to be
first order by the Mean Field approach. However it is well known that this
approach is not reliable insofar as the orders of the phase transitions are concerned.
(c) It should be pointed out that
there exists a phase separation on the vertical axis $\bt_g^\pr=0.$
This phase boundary is just an artifact of
the specific Mean Field approach that we have adopted. The problem has
already been recognised before \cite{reu} and alternative mean field approaches
have been devised, which avoid this unphysical behaviour. However they seem
rather ad hoc. In addition this phase transition is far from reality only
for very large values of $\bt_h$ (of order 1), while for smaller $\bt_h$
the phase transition is very near the real behaviour. Thus we have decided
not to use these alternatives.

\begin{figure}[!h]
\begin{center}
\includegraphics[scale=0.32]{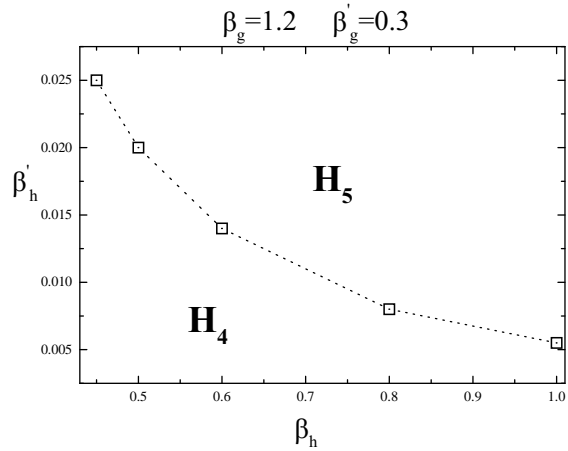}
\caption{Phase diagram in the $\bt_h - \bt_h^\pr$ plane for
$\bt_g=1.2,\bt^\pr_g=0.3,\beta_{R}=0.01.$}
\label{03}
\end{center}
\end{figure}

\begin{figure}[!h]
\begin{center}
\includegraphics[scale=0.32]{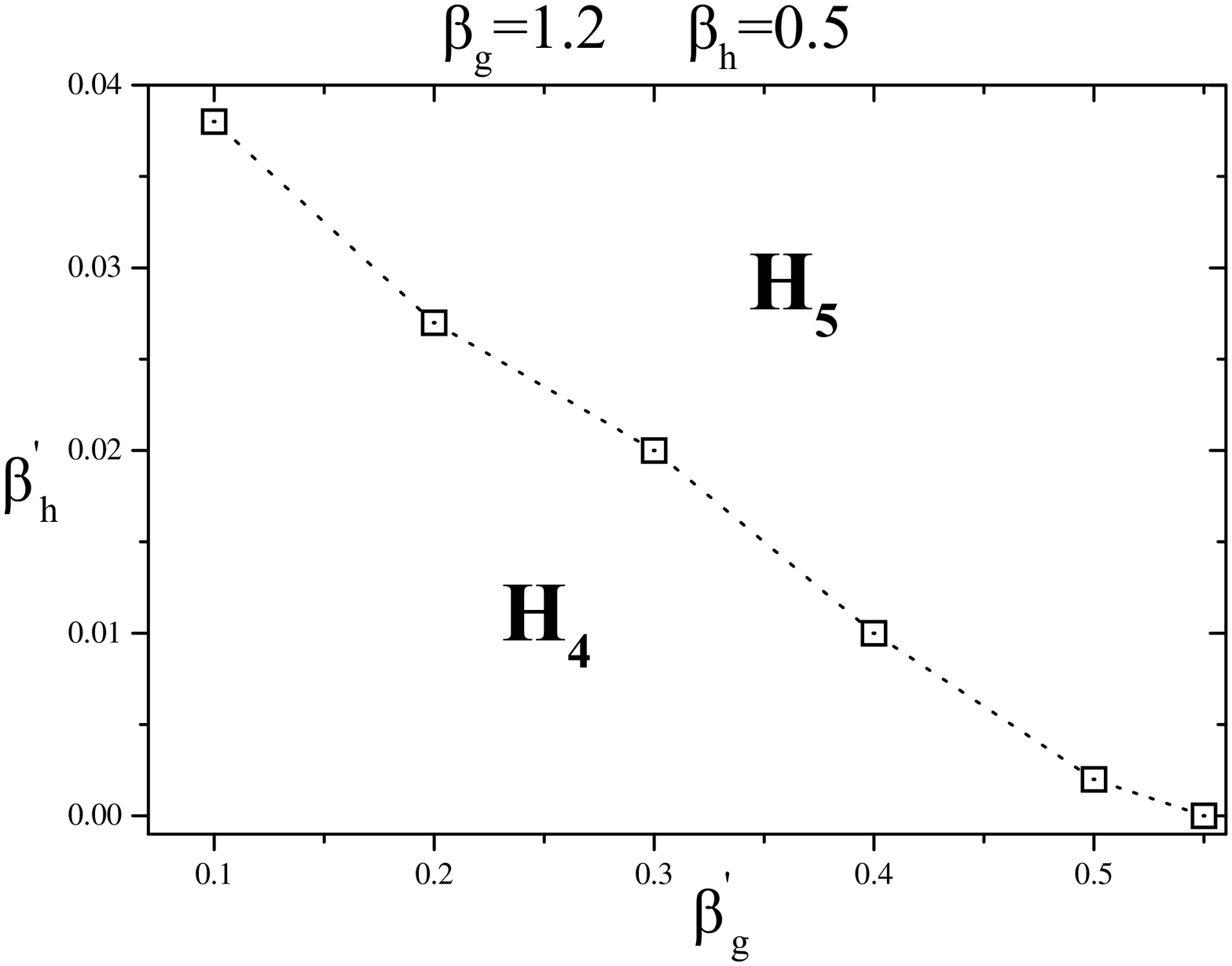}
\caption{Phase diagram in the $\bt_g^\pr - \bt_h^\pr$ plane for
$\bt_g=1.2,\bt_h=0.5,\beta_{R}=0.01.$}
\label{05}
\end{center}
\end{figure}

We may use the Mean Field approach to explore the role of the $\bt_h^\pr$
parameter. We expect that its role will be most important when the remaining
parameters initially locate the system in the $H_4$ phase. If we start with
$\bt_h^\pr=0$ and gradually increase its value, it is natural to expect that
the transverse direction will communicate more and more with the
space-like directions, so gradually the system will move from the $H_4$
to the $H_5$ phase. If the system lies initially in the $S$ or the $C$ phases,
it is conceivable that the system will move to a Higgs phase. However,
we have chosen to find the
critical $\bt_h^\pr$ for each value of $\bt_h (\bt_g^\pr)$ when
$\bt_g^\pr=0.3$ $(\bt_h=0.5)$ and determine in this way the critical line in the
$\bt_h - \bt_h^\pr$ plane ($\bt_g^\pr - \bt_h^\pr$ plane).
The results are contained in figures \ref{03} and \ref{05}. We see that if
$\bt_g^\pr$ (or $\bt_h$) are large enough, the critical $\bt_h^\pr$
tends to zero, that is the system is in the $H_5$ phase from the beginning.
In addition we see that the value $\bt_h^\pr=0.001$ that we have used
throughout the paper is small enough to permit for the existence of
an $H_4$ phase. Finally we note that we cannot detect here the phase
$C_4$ found by the Monte Carlo approach, namely the one with small
$P_T$ but large $L_T.$ This is again due to the specific structure of
our Mean Field approach. In particular the link consists of products of
the gauge link variable multiplied by products of the scalar field angular
variables (which yield bounded contributions). Thus it is not possible to
have small gauge variables on the links (yielding the small $P_T$) and at the
same time have a large $L_T.$

\begin{figure}[!h]
\begin{center}
\includegraphics[scale=0.32]{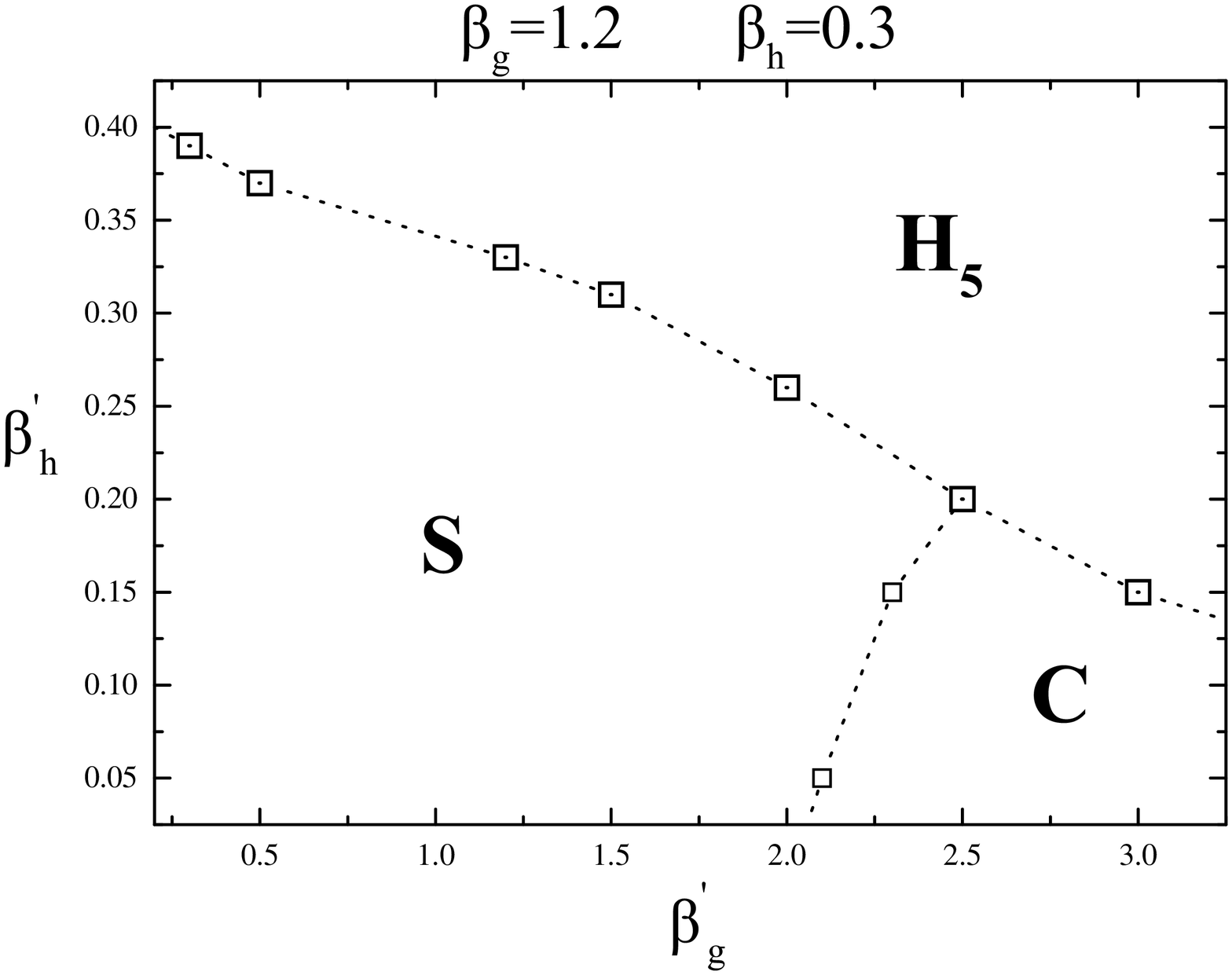}
\caption{Phase diagram in the $\bt_g^\pr - \bt_h^\pr$ plane for
$\bt_g=1.2,\bt_h=0.3,\beta_{R}=0.01.$}
\label{scl}
\end{center}
\end{figure}

In addition it is possible to start with $\bt_h^\pr=0$ and the remaining
variables such that the system is initially in the $S$ or the $C$ phases.
It turns out that increasing $\bt_h^\pr$ the system will undergo a phase
transition and move to the $H_5$ phase.
In figure \ref{scl} we have chosen $\bt_h=0.3, \bt_R=0.01, \bt_g=1.2$ and
scanned $\bt_h^\pr$ for various values of $\bt_g^\pr.$ In other words we
scanned various regions of the $S$ and $C$ phases. The result is
somehow more complicated than the ones in the previous two
figures. The system moves from either $S$ or $C$ to  $H_5.$

\end{document}